\documentclass[aip,amsmath,amssymb,reprint]{revtex4-2}
\usepackage[colorlinks,citecolor=blue]{hyperref}
\usepackage{graphicx}% Include figure files
\usepackage{dcolumn}% Align table columns on decimal point
\usepackage{bm}% bold math
\usepackage{esint} % for integrals
\usepackage{hyperref}
\begin{document}

\title[]{Unstable delayed feedback control to change sign of coupling strength for weakly coupled limit cycle oscillators}

\author{Viktor Novi\v{c}enko}
\email[]{viktor.novicenko@tfai.vu.lt}
\homepage[]{http://www.itpa.lt/~novicenko/}
\affiliation{Faculty of Physics, Vilnius University, Saul\.{e}tekio ave. 3, LT-10222 Vilnius, Lithuania}

\author{Irmantas Ratas}
\homepage[]{http://ndlab.lt/}
\affiliation{Center for Physical Sciences and Technology, Saul\.{e}tekio ave. 3, LT-10222 Vilnius, Lithuania}

\date{\today}% It is always \today, today,

\begin{abstract}
Weakly coupled limit cycle oscillators can be reduced into a system of weakly coupled phase models. These phase models are helpful to analyze the synchronization phenomena. For example, a phase model of two oscillators has a one-dimensional differential equation for the evolution of the phase difference. The existence of fixed points determines frequency-locking solutions. By treating each oscillator as a black-box possessing a single input and a single output, one can investigate various control algorithms to change the synchronization of the oscillators. In particular, we are interested in a delayed feedback control algorithm. Application of this algorithm to the oscillators after a subsequent phase reduction should give the same phase model as in the control-free case, but with a rescaled coupling strength. The conventional delayed feedback control is limited to the change of magnitude but does not allow the change of sign of the coupling strength. In this work, we present a modification of the delayed feedback algorithm supplemented by an additional unstable degree of freedom, which is able to change the sign of the coupling strength. Various numerical calculations performed with Landau-Stuart and FitzHugh-Nagumo oscillators show successful switching between an in-phase and anti-phase synchronization using the provided control algorithm. Additionally, we show that the control force becomes non-invasive if our objective is stabilization of an unstable phase difference for two coupled oscillators.
\end{abstract}

\maketitle

\begin{quotation}
More than 350 years ago, Huygens observed two pendulum clocks attached to a board placed on backs of two chairs. He discovered that, independently of the initial conditions of the pendulums, after a while, they became swinging toward each other and then apart. Nowadays, it is called anti-phase synchronization when two oscillators are coupled in such a way that the phases of the oscillators repulse each other, and the phase difference settles to $\pi$. Another spectacular synchronization case is when two metronomes are placed on a light platform that can roll on light cylinders. For such a case, both metronomes simultaneously swing to the same sides, exhibiting an in-phase synchronization when the coupling attracts one oscillator's phase to another. A theory behind the analysis of weakly coupled limit cycle oscillators, called a phase reduction, allows one to determine what kind of synchronous behavior may occur. In this work, we assume that one can measure some output from the oscillator, and based on that measurement, it is possible to affect the oscillator's state. This scheme is known as a black-box control with a single input and a single output. Under such a framework, we present the control algorithm capable of flipping the coupling sign in the phase equations. It means that under control, the oscillators ``feel'' opposite phase coupling than naturally exists. Thus, it potentially allows one to force the in-phase synchronization for the Huygen's clocks or the anti-phase synchronization for the metronomes. Moreover, the phase reduction predicts that if a control-free system possesses in-phase or anti-phase synchronization, the opposite synchronization also exists; yet, it is unstable. Once our control algorithm stabilizes the unstable synchronization point, the control force vanishes. Thus, the algorithm is non-invasive.
\end{quotation}

\section{\label{sec:intro}Introduction}

Starting from a famous Huygens' work on ``an odd kind sympathy'' synchronization as a phenomenon appears in various fields, such as biology~\cite{Buck1988,Friedrich12,Klindt2017}, chemistry~\cite{nakao14,Kiss2002}, physics~\cite{PhysRevE.57.1563,Weiss2016}, sociology~\cite{Neda2000,Neda2000a}, engineering systems~\cite{Motter2013,Dorfler2013,Pollakis2014}, etc. The main theoretical tool to investigate the synchronization between weakly coupled limit cycle oscillators is a phase reduction~\cite{pikov01,kura03,Nakao2016}. It allows us to write a dynamical equation only for the phase of a particular oscillator instead of dealing with a whole phase space.

The conventional phase reduction, which deals with ordinary differential equations, can be extended to delay differential equations~\cite{physd12,kot12}. The interesting results were obtained in Ref.~\cite{physd12} for the case of a delayed feedback control (DFC) force, which is constructed as a difference between delayed and undelayed feedback signals. It was shown that the phase reduction of an oscillator under the DFC gives the same phase equation as for the control-free case, only an inter-oscillatory coupling strength is re-scaled. This property can be exploited to ``effectively'' increase or decrease coupling strength and, as a consequence, to control the synchrony~\cite{Novicenko2015}. The universality of the re-scaling law allows one to implement the control of the synchronization independently on complexity of the particular oscillatory unit or topology of a network of coupled oscillators. A similar technique is used to achieve the in-phase synchronization for the near-identical oscillator units~\cite{Novicenko2018}; yet, the algorithm requires the knowledge of the topology of the network.

Despite the advantages mentioned above, the conventional delayed feedback scheme cannot change the sign of the coupling strength, and the reason for that is a so-called odd-number limitation theorem~\cite{hoo12} stating that the target limit cycle becomes unstable when the coupling strength flips its sign. The aim of this paper is to modify the delayed feedback controller by adding an additional unstable degree of freedom, such that the limit cycle is stable exactly in the case where the sign of the coupling strength is reversed.

The effect of the sign's flipping is illustrated on two coupled oscillators, which naturally are in a phase-locked regime. The phase difference equation of oscillators is one-dimensional and has stable and unstable fixed points representing in-phase and anti-phase synchronization states. The stability of the synchronous regime depends on a coupling form. The application of the controller to oscillators is equivalent to reversing of time flow; thus, it flips the stability of the synchronization points. Moreover, once the stabilized synchronization point is achieved, the control force vanishes, representing the non-invasive nature of the controller.

This paper is organized as follows. In Sec.~\ref{sec:prob}, we describe a mathematical formulation of the problem. Section~\ref{sec:udfc_oscillator} is devoted to the stability analysis of a single oscillator under the unstable delayed feedback control. Here we show that the stability almost always can be achieved. The main equations of the controller are presented by Eqs.~(\ref{eq:one_osc_udfcf}). In Sec.~\ref{sec:two_osc_in_anti}, the numerical results for the FitzHugh-Nagumo and the Landau-Stuart oscillators are presented. In Sec.~\ref{sec:non-inv}, we supplement the controller by including a slowly changing time delay and demonstrate the non-invasive nature of the controller. Section~\ref{sec:netw} demonstrates the disruption of a frequency-locking synchronization in the oscillator network by applying the controller to only one unit of the network. Finally, Sec.~\ref{sec:conc} contains a summary of the paper.

\section{\label{sec:prob}Problem formulation}

\subsection{\label{subs:weak}Complex network of weakly coupled limit cycle oscillators}

We study general class of $N$ limit cycle oscillators mutually coupled between each other and forming a network of an arbitrary topology. Each oscillator is assumed to be as a black-box under single-input single-output control,
\begin{subequations}
\label{eq:main}
\begin{align}
\dot{\mathbf{x}}^{(i)} &= \mathbf{f}^{(i)} \left(\mathbf{x}^{(i)} , r^{(i)} \right)+\varepsilon \sum_{\substack{j=1 \\ j\neq i}}^{N} a^{(ij)}\mathbf{G}^{(ij)}\left( \mathbf{x}^{(j)} , \mathbf{x}^{(i)}  \right), \label{eq:main_1} \\
s^{(i)}(t) &= g^{(i)} \left(\mathbf{x}^{(i)}(t) \right), \label{eq:main_2}
\end{align}
\end{subequations}
here, $\mathbf{x}^{(i)}$ represents the state vector of the $i$th oscillator, $\mathbf{f}^{(i)} (\mathbf{x},0 )$ is a vector field governing uncoupled and uncontrolled oscillators such that the differential equation ${\dot{\mathbf{x}}=\mathbf{f}^{(i)} (\mathbf{x},0 )}$ has a stable limit cycle solution ${\bm{\xi}^{(i)}(t+T^{(i)})=\bm{\xi}^{(i)}(t)}$ with the natural period $T^{(i)}$, the dimensionless coupling constant $\varepsilon$ is a small parameter of the system, $a^{(ij)}$ is an element of a network adjacency matrix that encodes network topology, and $\mathbf{G}^{(ij)}$ represents coupling function. In order to ensure unique factorization of the coupling term, we assume that $a^{(ij)}$ is equal either to one or zero. The scalar signal $s^{(i)}(t)$ represents single output, where the function $g^{(i)}(\mathbf{x}^{(i)})$ encodes transformation from the state vector to the measurable scalar quantity. The variable $r^{(i)}$ stands for single input and is constructed as a feedback force from the knowledge of $s^{(i)}$. We assume that the difference between the natural periods $|T^{(i)}-T^{(j)}|$ as well as the difference between the natural frequencies $|\Omega^{(i)}-\Omega^{(j)}|=|2\pi/T^{(i)}-2\pi/T^{(j)}|$ is of the same order as the small parameter $\varepsilon$. To be more precise, one can assume that the vector fields $\mathbf{f}^{(i)}$ parametrically depend on $\varepsilon$ such that all natural periods $T^{(i)}(\varepsilon)$ at the point $\varepsilon=0$ are equal to a ``central'' period $T=T^{(i)}(0)$. Our aim is to obtain analytical results as an expansion with respect to a small coupling constant, and throughout the paper, we restrict our analysis only to the zeroth and first-order terms of expansions with respect to $\varepsilon$. Thus, one can apply a combination of a phase reduction approach~\cite{kura03,pikov01,Nakao2016} and an averaging method~\cite{burd07}.

\subsection{\label{subs:phase_osc}Network of the phase-oscillators in a control-free regime}

In this subsection, we analyze the control-free regime, i.e. all inputs  $r^{(i)}=0$. For this case, one can temporarily ignore Eq.~(\ref{eq:main_2}). By picking a ``central'' frequency $\Omega=2\pi/T$ such that $|\Omega^{(i)}-\Omega|\sim \varepsilon$ and applying the phase reduction~\cite{kura03,pikov01} together with the averaging method~\cite{burd07}, the phase model of the oscillatory network~(\ref{eq:main_1}) in the rotating frame $\Omega$ reads
\begin{equation}
\dot{\psi}^{(i)} = \omega^{(i)}+\varepsilon \sum_{\substack{j=1 \\ j\neq i}}^{N} a^{(ij)} H^{(ij)}\left( \psi^{(j)}-\psi^{(i)} \right) ,
\label{eq:phase_cont_free}
\end{equation}
where $\psi^{(i)} \in [0, 2\pi)$ represents the phase of the $i$th oscillator, $\omega^{(i)}=\Omega^{(i)}-\Omega$ is a relative frequency, and a scalar phase-coupling function
\begin{equation}
\begin{aligned}
 H^{(ij)}(\chi) =& \frac{1}{T}\int\limits_{0}^{2\pi} \left[\mathbf{v}^{(i)}\left(\frac{s}{\Omega^{(i)}}\right) \right]^T \\ 
& \cdot \mathbf{G}^{(ij)}\left(\bm{\xi}^{(j)}\left(\frac{s+\chi}{\Omega^{(j)}}\right),\bm{\xi}^{(i)}\left(\frac{s}{\Omega^{(i)}}\right) \right)  \mathrm{d} s.
\label{eq:H_ij}
\end{aligned}
\end{equation}
Here, $[\cdot]^T$ denotes a transposed vector and $\mathbf{v}^{(i)}(t+T^{(i)})=\mathbf{v}^{(i)}(t)$ stands for a phase response curve (PRC) of the $i$th oscillator. Note that the PRC is a left Floquet mode corresponding to the trivial Floquet multiplier, $\mu=1$, and normalized to the right Floquet mode, $\dot{\bm{\xi}}^{(i)}(t)$, as $\int_0^{T_i} \left[\mathbf{v}^{(i)}(t) \right]^T \cdot \dot{\bm{\xi}}^{(i)}(t) \mathrm{d}t=1$.

\subsection{\label{subs:phase_osc_contr}Network of the phase-oscillators under delayed feedback control}

Now, let us analyze the complex network~(\ref{eq:main}) under the DFC force of the following form:
\begin{equation}
r^{(i)}(t) = K^{(i)} \left[ s^{(i)}\left(t-\tau^{(i)}\right)- s^{(i)}(t)\right],
\label{eq:r_dfc}
\end{equation}
where $\tau^{(i)}$ is a time delay of the $i$th control force and $K^{(i)}$ represents control gain. We assume that all time delays are close to the natural periods of the oscillators ${|\tau^{(i)}-T^{(i)}|\sim \varepsilon}$; therefore, $r^{(i)}(t)$ always remains small (${r^{(i)}(t)\sim \varepsilon}$). Such control force~(\ref{eq:r_dfc}) does not change the profile of the particular oscillator if the time delay is equal to the natural period, or in other words, for $\varepsilon=0$ and $\tau^{(i)}=T^{(i)}$, each oscillator has the same periodic solution as for the control-free case $\bm{\xi}^{(i)}(t)$. However, the stability of the limit cycle $\bm{\xi}^{(i)}(t)$ changes due to the control force, and, as a consequence, the oscillator's response to external perturbation changes too. As it is shown in Refs.~\cite{Novicenko2015,Novicenko2018}, the network~(\ref{eq:main}) together with~(\ref{eq:r_dfc}) reduces into the phase model similar to~(\ref{eq:phase_cont_free}),
\begin{equation}
\dot{\psi}^{(i)} = \omega^{(i)}_{\mathrm{eff}}+\varepsilon^{(i)}_{\mathrm{eff}} \sum_{\substack{j=1 \\ j\neq i}}^{N} a^{(ij)} H^{(ij)}\left( \psi^{(j)}-\psi^{(i)} \right).
\label{eq:phase_cont}
\end{equation}
The phase-coupling function $H^{(ij)}$ is not affected by the DFC, while the frequency and coupling constant change and take the following forms:
\begin{equation}
\varepsilon^{(i)}_{\mathrm{eff}} = \varepsilon \alpha\left( K^{(i)} C^{(i)} \right)
\label{eq:eps_eff}
\end{equation}
and
\begin{equation}
\omega^{(i)}_{\mathrm{eff}} = \omega^{(i)}+\Omega \frac{T^{(i)}-\tau^{(i)}}{T}\left[1-\alpha\left(K^{(i)} C^{(i)}\right) \right].
\label{eq:omega_eff}
\end{equation}
The function $\alpha$ has a simple form $\alpha(x)=(1+x)^{-1}$. The constant $C^{(i)}$ reads as an integral
\begin{equation}
\begin{aligned}
C^{(i)}=& \int\limits_0^{T^{(i)}} \left\lbrace \left[ \mathbf{v}^{(i)}(t)\right]^T\cdot D_2\mathbf{f}^{(i)}\left(\bm{\xi}^{(i)}(t),0 \right) \right\rbrace \\
& \cdot \left\lbrace \left[\nabla g^{(i)}\left(\bm{\xi}^{(i)}(t)\right)\right]^T \cdot \dot{\bm{\xi}}^{(i)}(t)\right\rbrace \mathrm{d}t.
\label{eq:h}
\end{aligned}
\end{equation}
Here, $D_2$ denotes differentiation of the vector field with respect to the second argument,
\begin{equation}
D_2\mathbf{f}\left(\bm{\xi}(t),0 \right)=\left.\frac{\partial \mathbf{f}\left(\bm{\xi}(t),r \right)}{\partial r}\right|_{r=0}.
\label{eq:D2}
\end{equation}

As one can see from (\ref{eq:eps_eff}) and (\ref{eq:omega_eff}), in the case of the time delays being equal to the natural periods $\tau^{(i)}=T^{(i)}$, the effective frequencies remain unaffected $\omega^{(i)}_{\mathrm{eff}}=\omega^{(i)}$, while the effective coupling strength rescales. Since the control force (\ref{eq:r_dfc}) may be constructed without knowledge of the particular oscillator's vector field, it can be used to control synchronization in the network~\cite{Novicenko2015}. By appropriate choice of the control gain $K^{(i)}$, the re-scaling factor $\alpha$ may gain any positive or negative values. In particular, for $\left(K^{(i)} C^{(i)}\right) < -1$, the factor $\alpha$ is negative and the effective coupling strength $\varepsilon^{(i)}_{\mathrm{eff}}$ changes its sign. Nevertheless, the problem here is that the phase model~(\ref{eq:phase_cont}) is relevant only until the limit cycle $\bm{\xi}^{(i)}(t)$ is stable. According to the odd-number limitation theorem~\cite{hoo12}, the periodic orbit $\bm{\xi}^{(i)}(t)$ is an unstable solution of the differential equation $\dot{\mathbf{x}}^{(i)} = \mathbf{f}^{(i)} \left(\mathbf{x}^{(i)}, r^{(i)} \right)$ with $r^{(i)}(t)$ being of the form of~(\ref{eq:r_dfc}) and  with $\tau^{(i)}=T^{(i)}$ if the DFC control gain satisfies the inequality
\begin{equation}
(-1)^m \left(1+K^{(i)} C^{(i)}\right) < 0,
\label{eq:odd}
\end{equation}
where $m$ is the number of real Floquet multipliers (FMs) larger than $1$ existing in the control-free system $\dot{\mathbf{x}}^{(i)} = \mathbf{f}^{(i)} \left(\mathbf{x}^{(i)}, 0 \right)$. By definition, the limit cycle $\bm{\xi}^{(i)}(t)$ is stable; therefore, $m=0$ and the inequality~(\ref{eq:odd}) reduces to $\left(K^{(i)} C^{(i)} \right)<-1$. In other words, the DFC force made the periodic orbit $\bm{\xi}^{(i)}(t)$ unstable exactly at the point where $\varepsilon^{(i)}_{\mathrm{eff}}$ flips its sign. Note that the odd-number limitation theorem does not guarantee stability of $\bm{\xi}^{(i)}(t)$ for the control gain value that violates~(\ref{eq:odd}); therefore, the violation of~(\ref{eq:odd}) is only necessary but not sufficient condition for the stability of the periodic solution $\bm{\xi}^{(i)}(t)$.

The main goal of this work is to present the control algorithm based on the form of the DFC, which allows one to set negative values of $\alpha$ while preserving the stability of the limit cycle. As a consequence, the phase model~(\ref{eq:phase_cont}) becomes valid having an opposite sign of $\varepsilon^{(i)}_{\mathrm{eff}}$ in comparison with the natural coupling constant $\varepsilon$. The main idea is based on Ref.~\cite{pyr01} and schematically may be explained by the following steps. We add an additional unstable degree of freedom and couple it with the oscillator in such a way that the periodic solution does not change the profile $\bm{\xi}^{(i)}(t)$ but make it unstable. Then we add the DFC force that stabilizes the periodic solution. Such stabilization is achieved for the values of the control gain $K^{(i)}$, which give $\alpha<0$. The ``bypass'' of the odd-number limitation theorem is achieved since the additional degree of freedom adds a real Floquet multiplier with the value being larger than one, and as a consequence, the theorem~(\ref{eq:odd}) gives an opposite outcome; i.e., the limit cycle is unstable for $\left(K^{(i)} C^{(i)}\right)>-1$ and may be stable only for $\left(K^{(i)} C^{(i)}\right)<-1$.

\section{\label{sec:udfc_oscillator}One oscillator under delayed feedback control supplemented by an unstable degree of freedom}

The results presented in Sec.~\ref{sec:prob} are valid for any stable limit cycle oscillators, governed by $\mathbf{f}^{(i)}$, even with a different dimension for different indexes $i$, and the only main requirement is that $\left|T^{(i)}-T^{(j)}\right|\sim \left|\tau^{(i)}-T^{(i)}\right| \sim \varepsilon$ should be small. In order to implement the goal formulated in Sec.~\ref{subs:phase_osc_contr}, we can study only one particular oscillator supplemented by an additional unstable degree of freedom and affected by the DFC. Thus, in this section, we will drop the superscript $i$ for all quantities as being the notation for different oscillators and instead use it for different purposes.

By adding an additional degree of freedom, the previous single-input single-output problem becomes the two-inputs two-outputs problem and the control gain $K$ becomes matrix $\mathbf{K}$ of the form of 2$\times$2. To deal with it, we will use a similar technique as in Ref.~\cite{Pyragas2013}; i.e., we factorize the control matrix $\mathbf{K}=\kappa \tilde{\mathbf{K}}$ into a scalar dimensionless control gain $\kappa$ and a control form $\tilde{\mathbf{K}}$. To make the factorization uniquely defined, we assume that the elements of $\tilde{\mathbf{K}}$ are in the interval $\tilde{K}_{ij} \in [-1, 1]$ and at least one of the elements is equal to $-1$ or $1$. The scalar control gain $\kappa$ is assumed to be positive, $\kappa>0$, since $\kappa<0$ can be covered by reverting the sign of $\tilde{\mathbf{K}}$. The equations governing oscillator's dynamics read
\begin{subequations}
\label{eq:one_osc_udfc}
\begin{align}
\dot{\mathbf{x}} &= \mathbf{f} \left(\mathbf{x}, r \right), \label{eq:one_osc_udfc_1} \\
\dot{w} &= \left( \lambda_{\mathrm{l}}+\lambda_{\mathrm{n}} s(t) \right) w \nonumber \\
& + \kappa \left\lbrace \tilde{K}_{21} \left[ s(t-T)-s(t) \right] + \tilde{K}_{22} \left[ w(t-T)-w(t) \right] \right\rbrace, \label{eq:one_osc_udfc_2} \\
s(t) &= g \left(\mathbf{x}(t) \right), \label{eq:one_osc_udfc_3} \\
r(t)&= w(t) \nonumber \\
& +\kappa \left\lbrace \tilde{K}_{11} \left[ s(t-T)-s(t) \right] + \tilde{K}_{12} \left[ w(t-T)-w(t) \right] \right\rbrace \label{eq:one_osc_udfc_4} ,
\end{align}
\end{subequations}
here, $\mathbf{x}$ stands for an $n$-dimensional state vector, $w(t)$ is a new dynamical variable representing an unstable degree of freedom, and $\lambda_{\mathrm{l}}$ and $\lambda_{\mathrm{n}}$ are linear and non-linear contributions to the dynamics of $w(t)$ respectively.

Without the DFC ($\kappa=0$), the $(n+1)$-dimensional system~(\ref{eq:one_osc_udfc}) possesses the periodic solution $(\mathbf{x}^T (t),w(t))=(\bm{\xi}^T (t+T),0)=(\bm{\xi}^T (t),0)$. As it is shown in Appendix~\ref{appsec:fm_and_prc}, such a solution has the first $n$ FMs and 
the first $n$ components of the PRC the same as in the case of an $n$-dimensional stable oscillator without $w(t)$. The last FM
\begin{equation}
\mu_{n+1}=\exp\left[ \lambda_{\mathrm{l}}T+\lambda_{\mathrm{n}} \int_0^T g \left(\bm{\xi}(t) \right) \mathrm{d}t \right]
\label{eq:fm_n1}
\end{equation}
and the corresponding Floquet exponent (FE) $\Lambda_{n+1}=\lambda_{\mathrm{l}}+\lambda_{\mathrm{n}} T^{-1} \int_0^T g \left(\bm{\xi}(t) \right) \mathrm{d}t$.

\begin{figure}[h!]
\centering\includegraphics[width=0.95\columnwidth]{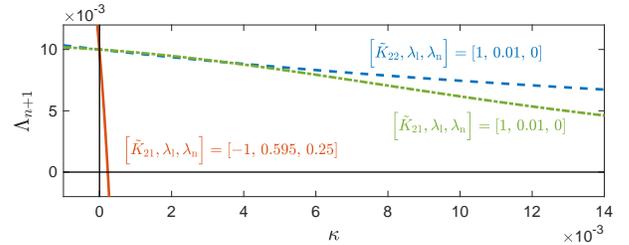}
\caption{\label{fig:FE_FHN_DFC} Three examples of the dependence of the unstable FE vs the control gain for the FitzHugh-Nagumo oscillator model. The model is described further in Sec.~\ref{subsec:two_fhn} by Eqs.~(\ref{fhn_f}), (\ref{fhn_s}). In all cases the matrix $\tilde{\mathbf{K}}$ has all elements equal to zero except the element tagged in the plot. The starting position of the unstable FE $\Lambda_{n+1}(\kappa=0)=0.01$ is assumed to be small. However, only the solid orange line shows stabilization for the small $\kappa$, while the dashed blue and dashed-dotted green lines remain unstable.}
\end{figure}

The change of the control gain $\kappa$ from zero leads to movement of FMs. The aim of changing $\kappa$ is to push the $\mu_{n+1}$ into the unit circle through trivial FM, while other FMs remain in this circle. As a consequence, the system~(\ref{eq:one_osc_udfc}) will become stable.  By appropriate choice of $\lambda_{\mathrm{l}}$ and $\lambda_{\mathrm{n}}$, it may be beneficial to set $\mu_{n+1}$ very close to trivial FM; i.e., $\mu_{n+1} \rightarrow 1+0$ and $\Lambda_{n+1} \rightarrow +0$. The $\mu_{n+1}$ closeness to trivial FM would allow one to expect that it enters a unit circle before other FMs go out of it. However, such a strategy may have some difficulties as it is shown in the examples depicted in Fig.~\ref{fig:FE_FHN_DFC}.

When the parameters $\lambda_{\mathrm{l}}$, $\lambda_{\mathrm{n}}$, and $\tilde{\mathbf{K}}$ are set, the function $\Lambda_{n+1}(\kappa)$ is well determined, and the derivative $\left.\mathrm{d}\Lambda_{n+1}(\kappa)/\mathrm{d}\kappa \right|_{\kappa=0}$ indicates where does unstable FE move by slightly increasing $\kappa$. From Fig.~\ref{fig:FE_FHN_DFC} one can see that in all three cases, the derivative $\left.\mathrm{d}\Lambda_{n+1}(\kappa)/\mathrm{d}\kappa \right|_{\kappa=0}$ is negative; therefore, one can think that by setting-up small enough value of the unstable FE, $\Lambda_{n+1}(0)$, we can always achieve the stabilization. Unfortunately, that is not true due to the fact that the derivative $\left.\mathrm{d}\Lambda_{n+1}(\kappa)/\mathrm{d}\kappa \right|_{\kappa=0}$ depends on $\lambda_{\mathrm{l}}$ and $\lambda_{\mathrm{n}}$ and as a consequence on the initial value of the unstable FE, $\Lambda_{n+1}(0)$. Since we assume that $\Lambda_{n+1}(0)$ is small, one can perform the expansion of the derivative $\left.\mathrm{d}\Lambda_{n+1}(\kappa)/\mathrm{d}\kappa \right|_{\kappa=0}$ in terms of $\Lambda_{n+1}(0)$. In two dimensional parameter space $(\lambda_{\mathrm{l}};\lambda_{\mathrm{n}})$, there is a straight line giving $\Lambda_{n+1}(0)=0$; thus, along the perpendicular direction of the line, one can write
\begin{equation}
\left.\frac{\mathrm{d}\Lambda_{n+1}(\kappa)}{\mathrm{d}\kappa} \right|_{\kappa=0} = c_0+c_1 \Lambda_{n+1}(0)+c_2 \Lambda_{n+1}^2(0)+\cdots.
\label{eq:ex_der}
\end{equation}
Typically, the constant $c_0=0$ (that is the case for the dashed blue and dashed-dotted green lines of Fig.~\ref{fig:FE_FHN_DFC}); thus,
\begin{equation}
\left.\frac{\mathrm{d}\Lambda_{n+1}(\kappa)}{\mathrm{d}\kappa} \right|_{\kappa=0} \sim \mathcal{O}(\Lambda_{n+1}(0)).
\label{eq:dfe_n1}
\end{equation}
As a consequence, the closer we put the $\Lambda_{n+1}(0)$ to zero, the slower it moves as $\kappa$ increases from zero. On the other hand, for $c_0\neq 0$, the system~(\ref{eq:one_osc_udfc}) becomes stable for the small value of the control gain, $\kappa \sim \mathcal{O}(\Lambda_{n+1}(0))$. Further in this section, we will show that there are two essential requirements for the control parameters, namely,

\hypertarget{cond1}{(i)} The control form $\tilde{\mathbf{K}}$ has non-zero element $\tilde{K}_{21}$, while other elements may be equal to zero.

\hypertarget{cond2}{(ii)} The dynamics of $w(t)$  has a non-zero non-linear part; $\lambda_{\mathrm{n}} \neq 0$.
\\*
The fulfillment of the conditions \hyperlink{cond1}{(i)} and \hyperlink{cond2}{(ii)} (as, for example, the solid orange line in Fig.~\ref{fig:FE_FHN_DFC}) is necessary and sufficient in order to achieve stabilization in the presence of $\Lambda_{n+1}(0) \rightarrow +0$. Note that it does not mean that other matrix $\tilde{\mathbf{K}}$ coefficients are useless in a practical implementation of the control algorithm. As it is shown in Sec.~\ref{subsec:summ_lim}, the limit $\Lambda_{n+1}(0) \rightarrow +0$ might lead to some difficulties. Thus, the only aim of Sec.~\ref{sec:udfc_oscillator} is to show that theoretically, one can always stabilize the system~(\ref{eq:one_osc_udfc}) by employing \hyperlink{cond1}{(i)} and \hyperlink{cond2}{(ii)}. However, in practical implementation, the requirements \hyperlink{cond1}{(i)} and \hyperlink{cond2}{(ii)}  might be a good starting point and, in principle, can be relaxed.

The logical structure of this section is organized as follows: in Secs.~\ref{subsec:dfc_pfc} and~\ref{subsec:triv_fe} we introduce general formalism related to a proportional feedback and the delay feedback control schemes without involving smallness of $\Lambda_{n+1}(0)$. In Sec.~\ref{subsec:triv_fe_lim}, we reconsider the results of Sec.~\ref{subsec:triv_fe} in the presence of $\Lambda_{n+1}(0) \rightarrow +0$ by showing that the conditions \hyperlink{cond1}{(i)} and \hyperlink{cond2}{(ii)} are necessary for the stabilization. In Sec.~\ref{subsec:uns_fe}, we present the sufficiency of the conditions \hyperlink{cond1}{(i)} and \hyperlink{cond2}{(ii)}, and Sec.~\ref{subsec:summ_lim} is devoted to a summary and a discussion on the limitations.

\subsection{\label{subsec:dfc_pfc}Relation between Floquet exponents and Floquet modes for a system under proportional feedback and delayed feedback control }

Next, let us assume that we have fixed control form $\tilde{\mathbf{K}}$ and we increase the control gain $\kappa$. By changing $\kappa$, the FMs are moving. The information on the dynamics of the FMs in a delayed feedback controlled system~(\ref{eq:one_osc_udfc}) can be extracted from a similar system controlled by the proportional feedback~\cite{pyr02}. The analog of the system~(\ref{eq:one_osc_udfc}), where the DFC is replaced by proportional feedback control~(PFC), reads
\begin{subequations}
\label{eq:one_osc_pfc}
\begin{align}
\dot{\mathbf{x}} &= \mathbf{f} \left(\mathbf{x}, w+\Gamma \left\lbrace \tilde{K}_{11} \left[ g(\bm{\xi}(t))-s(t) \right] + \tilde{K}_{12} \left[-w(t) \right] \right\rbrace \right), \label{eq:one_osc_pfc_1} \\
\dot{w} &= \left( \lambda_{\mathrm{l}}+\lambda_{\mathrm{n}} s(t) \right) w \nonumber \\
& + \Gamma \left\lbrace \tilde{K}_{21} \left[ g(\bm{\xi}(t))-s(t) \right] + \tilde{K}_{22} \left[ -w(t) \right] \right\rbrace, \label{eq:one_osc_pfc_2} \\
s(t) &= g \left(\mathbf{x}(t) \right). \label{eq:one_osc_pfc_3}
\end{align}
\end{subequations}
Here, $\Gamma$ is the dimensionless PFC control gain. By introducing notations for $n\times n$ Jacobian matrix
\begin{equation}
\mathbf{A}(t)=D_1 \mathbf{f}\left( \bm{\xi}(t) ,0 \right),
\label{eq:jac}
\end{equation}
$n \times 1$ vector of the derivatives with respect to the input signal
\begin{equation}
\mathbf{p}(t)=D_2 \mathbf{f}\left( \bm{\xi}(t) ,0 \right)
\label{eq:p}
\end{equation}
and $n \times 1$ vector of the derivatives with respect to the dynamical variables
\begin{equation}
\mathbf{q}(t)= \nabla g( \bm{\xi}(t)),
\label{eq:q}
\end{equation}
one can write the evolution of small perturbation from the limit cycle for both PFC system~(\ref{eq:one_osc_pfc}) and DFC system~(\ref{eq:one_osc_udfc}) as
\begin{equation}
\label{eq:delta_pfc_dfc}
\begin{aligned}
\left(
\begin{array}{c}
\delta \dot{\mathbf{x}} \\
\delta \dot{w}
\end{array}
  \right) &= \left(
  \begin{array}{rl}
  \mathbf{A}(t) & \mathbf{p}(t) \\
  \mathbf{0}^T_n & \lambda_{\mathrm{l}}+\lambda_{\mathrm{n}} g(\bm{\xi}(t))
  \end{array}
     \right)
     \left(
\begin{array}{c}
\delta \mathbf{x} \\
\delta w
\end{array} \right)  \\
& - \left\lbrace 
\begin{array}{c}
\Gamma, \;\mathrm{for}\,\mathrm{PFC} \\
\kappa, \;\mathrm{for}\,\mathrm{DFC}
\end{array}
\right\rbrace
\left(
\begin{array}{rl}
 \tilde{K}_{11} \mathbf{p}(t) \mathbf{q}^T(t) &  \tilde{K}_{12} \mathbf{p}(t) \\
 \tilde{K}_{21} \mathbf{q}^T(t) &  \tilde{K}_{22}
\end{array}
\right)
 \\
& \cdot \left[ \left(
\begin{array}{c}
\delta \mathbf{x} \\
\delta w
\end{array} \right)-
\left\lbrace 
\begin{array}{c}
0, \;\mathrm{for}\,\mathrm{PFC} \\
1, \;\mathrm{for}\,\mathrm{DFC}
\end{array}
\right\rbrace
\left(
\begin{array}{c}
\delta \mathbf{x}(t-T) \\
\delta w(t-T)
\end{array} \right)
 \right] .
\end{aligned}
\end{equation}
Here, $\mathbf{0}_n$ is an $n$-dimensional column-vector filled with zeros. Let us denote FEs of the DFC and PFC systems as $\Lambda_{\mathrm{D}}(\kappa)$ and $\Lambda_{\mathrm{P}}(\Gamma)$ and the corresponding right Floquet modes as $\mathbf{u}_{\mathrm{D}}(t,\kappa)$ and $\mathbf{u}_{\mathrm{P}}(t,\Gamma)$, respectively. The substitution of the form $(\delta\mathbf{x}^T (t),\delta w(t))=\exp\left( \Lambda_{\left\lbrace \mathrm{P}, \mathrm{D}\right\rbrace} t \right) \mathbf{u}^T_{\left\lbrace \mathrm{P},\mathrm{D} \right\rbrace}(t,\left\lbrace \Gamma,\kappa \right\rbrace)$ to Eq.~(\ref{eq:delta_pfc_dfc}) gives
\begin{equation}
\label{eq:fm_pfc_dfc}
\begin{aligned}
& \dot{\mathbf{u}}_{X}+\Lambda_{X} \mathbf{u}_{X} = \left(
  \begin{array}{rl}
  \mathbf{A}(t) & \mathbf{p}(t) \\
  \mathbf{0}^T_n & \lambda_{\mathrm{l}}+\lambda_{\mathrm{n}} g(\bm{\xi}(t))
  \end{array}
     \right)
     \mathbf{u}_{X} \\
& - \left\lbrace 
\begin{array}{r}
\Gamma, \, \mathrm{for} \, X=\mathrm{P} \\
\kappa\left(1-\mathrm{e}^{-\Lambda_{X} T}\right), \, \mathrm{for} \, X=\mathrm{D}
\end{array}
\right\rbrace \\
& \cdot \left(
\begin{array}{rl}
 \tilde{K}_{11} \mathbf{p}(t) \mathbf{q}^T(t) &  \tilde{K}_{12} \mathbf{p}(t) \\
 \tilde{K}_{21} \mathbf{q}^T(t) &  \tilde{K}_{22}
\end{array}
\right)
\mathbf{u}_{X} , \; \mathrm{with} \, X \in \left\lbrace \mathrm{P}, \mathrm{D} \right\rbrace.
\end{aligned}
\end{equation}
If both FEs are real (do not have imaginary parts), then one can obtain following relations between Floquet modes, FEs, and control gains of both systems:
\begin{subequations}
\label{eq:dp_relation}
\begin{align}
\mathbf{u}_{\mathrm{D}}(t,\kappa(\Gamma)) &=\mathbf{u}_{\mathrm{P}}(t,\Gamma), \label{eq:dp_relation_1} \\
\Lambda_{\mathrm{D}}(\kappa(\Gamma)) &=\Lambda_{\mathrm{P}}(\Gamma), \label{eq:dp_relation_2} \\
\kappa(\Gamma) &= \frac{\Gamma}{1-\exp\left( -\Lambda_{\mathrm{P}}(\Gamma) T \right)}. \label{eq:dp_relation_3}
\end{align}
\end{subequations}
The last equations allow us to map the FE of the PFC system into the FE in the DFC system and vice versa.

Note that Eq.~(\ref{eq:dp_relation}) works only for the real FEs, or in other words for the positive FMs, $\mu>0$. For the negative FMs, $\mu<0$, the FE is complex and has real part $\Re\left(\Lambda\right)=\ln|\mu|/T$ and imaginary part $\Im\left(\Lambda\right)=\mathrm{i}(\pi+2\pi k)/T$ with any integer $k$; therefore, the relations are similar to~(\ref{eq:dp_relation}). The only difference is that (\ref{eq:dp_relation_3}) becomes $\kappa(\Gamma)=\Gamma/ \left[1+\exp(-\Re(\Lambda_{\mathrm{P}}(\Gamma))T) \right]$. However, for the pair of complex conjugate FMs, the relations between the DFC and PFC systems are difficult to derive.

\subsection{\label{subsec:triv_fe}Trivial Floquet exponent}

The DFC system is autonomous; thus, for all $\kappa$, it has trivial Floquet exponent $\Lambda_{\mathrm{D},1}(\kappa)=0$ (here we use subscript $(\cdot)_1$ to denote the first FE). By rewriting~(\ref{eq:dp_relation_3}) as $\Gamma(\kappa)=\kappa\left[1-\exp(-\Lambda_{\mathrm{D}}(\kappa)T) \right]$, one can see that the trivial branch $\Lambda_{\mathrm{D},1}(\kappa)$ of the DFC system maps into one point $\Lambda_{\mathrm{P},1}(\Gamma=0)=0$ of the PFC system. For ${\Gamma\neq 0}$, the PFC system is non-autonomous; thus, in general, $\Lambda_{\mathrm{P},1}(\Gamma)\neq 0$. By mapping the trivial branch $\Lambda_{\mathrm{P},1}(\Gamma)$ segment near the point $\Gamma = 0$ into the DFC system, one can extract information on another (non-trivial) branch of the FE, which crosses the stability point, $\Lambda_{\mathrm{D}}=0$. Indeed, for $\Gamma \rightarrow 0$,  the first FE $\Lambda_{\mathrm{P},1}(\Gamma)\rightarrow0$; therefore, Eq.~(\ref{eq:dp_relation_3}) might give a finite value
\begin{equation}
\label{eq:kappa_tr}
\kappa^{*}=\lim_{\Gamma \rightarrow 0} \kappa(\Gamma)
\end{equation}
corresponding to a threshold control gain where some branch $\Lambda_{\mathrm{D}}(\kappa)$ crosses zero; i.e., $\Lambda_{\mathrm{D}}(\kappa^*)=0$.

\begin{figure}[h!]
\centering\includegraphics[width=0.95\columnwidth]{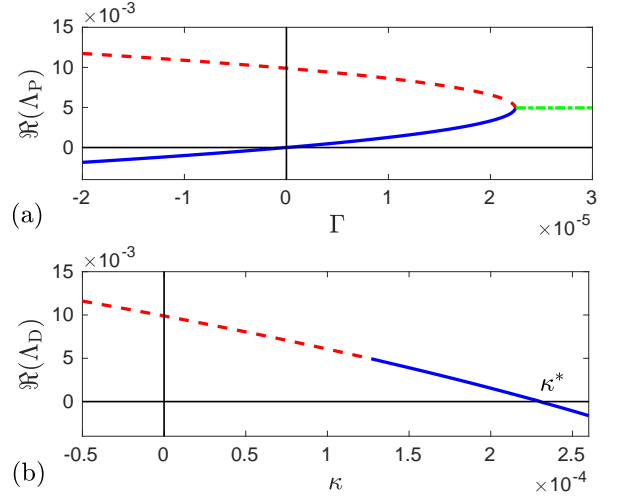}
\caption{\label{fig:pfc_dfc} The dependence of the FEs on the control gain in the case of PFC (panel (a)) and DFC (panel (b)) for the FitzHugh-Nagumo oscillator model. The control form $\tilde{\mathbf{K}}$ has non-zero element $K_{21}=-1$ and the linear and non-linear contributions to the dynamics of the unstable degree of freedom are $\lambda_{\mathrm{l}}=0.595$ and $\lambda_{\mathrm{n}}=-0.25$. Blue solid and red dashed lines in panel~(a) represent the trivial FE and the unstable FE, respectively. Both branches merge and a pair of complex conjugate FEs appear, which is depicted by a green dashed-dotted line. The unstable FE of panel (b) was obtained by  mapping the trivial and unstable branches of the PFC system to the DFC system.}
\end{figure}

Let us analyze a particular example, FitzHugh-Nagumo oscillator model described in Sec.~\ref{subsec:two_fhn} by Eqs.~(\ref{fhn_f}), (\ref{fhn_s}). In Fig.~\ref{fig:pfc_dfc}, the FEs of the PFC (panel (a)) and DFC (panel (b)) systems are depicted. The parameters $\tilde{\mathbf{K}}$, $\lambda_{\mathrm{l}}$, and $\lambda_{\mathrm{n}}$ are chosen in such a way that \hyperlink{cond1}{(i)} and \hyperlink{cond2}{(ii)} constraints are satisfied (for exact values, see the description of the figure). The trivial $\Lambda_{\mathrm{P},1}(\Gamma)$ and the unstable $\Lambda_{\mathrm{P},n+1}(\Gamma)$ FE branches of the PFC system using~(\ref{eq:dp_relation}) are mapped to one unstable FE branch $\Lambda_{\mathrm{D},n+1}(\kappa)$ of the DFC system.

The point $\kappa^{*}$ where $\Lambda_{\mathrm{D},n+1}(\kappa)$ crosses the zero can be estimated from the expansion of the trivial branch $\Lambda_{\mathrm{P},1}(\Gamma)$ and the corresponding right Floquet mode $\mathbf{u}_{\mathrm{P},1}(t,\Gamma)$ near $\Gamma=0$,
\begin{equation}
\label{eq:l_tr}
\Lambda_{\mathrm{P},1}(\Gamma)=0+\Lambda^{\prime}_{\mathrm{P},1}(0)\Gamma+\mathcal{O}(\Gamma^2),
\end{equation}
\begin{equation}
\label{eq:fm_tr}
\mathbf{u}_{\mathrm{P},1}(t,\Gamma)=\mathbf{u}^{(0)}_{\mathrm{P},1}(t)+\mathbf{u}^{(1)}_{\mathrm{P},1}(t)\Gamma+\mathcal{O}(\Gamma^2).
\end{equation}
Note that further, for the sake of simplicity, we will skip subscript $(\cdot)_{\mathrm{P}}$ and superscript $(\cdot)^{(0)}$ for the zeroth-order expansion term of the Floquet modes $\mathbf{u}^{(0)}_{\mathrm{P},i}(t)\equiv \mathbf{u}_{i}(t)$. The zeroth-order term is known analytically, $\mathbf{u}_{1}(t)=(\dot{\bm{\xi}}^T(t),0)^T$. By substituting~(\ref{eq:l_tr}) and (\ref{eq:fm_tr}) into (\ref{eq:fm_pfc_dfc}) and collecting $\mathcal{O}(\Gamma)$ order terms, we get
\begin{equation}
\label{eq:fm_pfc_first}
\begin{aligned}
 &\dot{\mathbf{u}}^{(1)}_{\mathrm{P},1} (t) = \left(
  \begin{array}{rl}
  \mathbf{A}(t) & \mathbf{p}(t) \\
  \mathbf{0}^T_n & \lambda_{\mathrm{l}}+\lambda_{\mathrm{n}} g(\bm{\xi}(t))
  \end{array}
     \right)
     \mathbf{u}^{(1)}_{\mathrm{P},1} (t)\\
& - \left[  \left(
\begin{array}{rl}
 \tilde{K}_{11} \mathbf{p}(t) \mathbf{q}^T(t) &  \tilde{K}_{12} \mathbf{p}(t) \\
 \tilde{K}_{21} \mathbf{q}^T(t) &  \tilde{K}_{22}
\end{array}
\right) +\Lambda^{\prime}_{\mathrm{P},1}(0) \mathbf{I}_{n+1} \right]
\left(
\begin{array}{c}
\dot{\bm{\xi}}(t) \\
0
\end{array}
\right),
\end{aligned}
\end{equation}
where $\mathbf{I}_{n+1}$ stands for $(n+1)\times(n+1)$ identity matrix. The last equation allows us to find the expansion coefficient $\Lambda^{\prime}_{\mathrm{P},1}(0)$. To do so, one should do the following three steps: first, multiply Eq.~(\ref{eq:fm_pfc_first}) from the left hand side by the PRC $\mathbf{v}_1^T(t)$ (note that the PRC is the first left Floquet mode, and the left and right Floquet modes are bi-orthogonal to each other); second, multiply differential equation for the PRC~(\ref{eqapp:prc}) by $\mathbf{u}^{(1)}_{\mathrm{P},1} (t)$ from the right hand side; and third, sum results of the first and the second steps. As a result, one gets
\begin{equation}
\label{eq:for_a}
\begin{aligned}
&\frac{\mathrm{d}}{\mathrm{d} t}\left(\mathbf{v}_1^T (t) \cdot \mathbf{u}^{(1)}_{\mathrm{P},1} (t) \right) =  - \Lambda^{\prime}_{\mathrm{P},1}(0) \\
& -\mathbf{v}_1^T (t)  \left(
\begin{array}{rl}
 \tilde{K}_{11} \mathbf{p}(t) \mathbf{q}^T(t) &  \tilde{K}_{12} \mathbf{p}(t) \\
 \tilde{K}_{21} \mathbf{q}^T(t) &  \tilde{K}_{22}
\end{array}
\right)  \left(
\begin{array}{c}
\dot{\bm{\xi}}(t) \\
0
\end{array}
\right).
\end{aligned}
\end{equation}
By integrating the last equation on the interval $[0,T]$, one can see that the derivative $\Lambda^{\prime}_{\mathrm{P},1}(0)$ can be written as a linear combination of the control matrix coefficients $\tilde{K}_{ij}$ (for further simplifications, we multiply the derivative by the period)
\begin{equation}
\label{eq:a}
\Lambda^{\prime}_{\mathrm{P},1}(0) T=-\left( \tilde{K}_{11} C_{11}+\tilde{K}_{12} C_{12}+\tilde{K}_{21} C_{21}+\tilde{K}_{22} C_{22} \right),
\end{equation}
where
\begin{subequations}
\label{eq:for_c}
\begin{align}
C_{11} &= \int\limits_0^{T} \mathbf{v}_{1,1:n}^T (t)  \mathbf{p}(t) \mathbf{q}^T(t) \dot{\bm{\xi}}(t) \mathrm{d}t, \label{eq:for_c_1} \\
C_{12} &= 0, \label{eq:for_c_2} \\
C_{21} &= \int\limits_0^{T} v_{1,n+1} (t) \mathbf{q}^T(t) \dot{\bm{\xi}}(t) \mathrm{d}t, \label{eq:for_c_3} \\
C_{22} &= 0. \label{eq:for_c_4}
\end{align}
\end{subequations}
Here $\mathbf{v}_{1,1:n}(t)$ denotes the vector constructed from the vector $\mathbf{v}_{1}(t)$ by dropping the $(n+1)$th component. Note that the coefficient (\ref{eq:for_c_1}) is exactly the same coefficient $C^{(i)}$ presented in Sec.~\ref{subs:phase_osc_contr} (cf. the definition (\ref{eq:h})). The derivative $\Lambda^{\prime}_{\mathrm{P},1}(0) T$ is inverse to $\kappa^{*}$. Indeed substituting~(\ref{eq:l_tr}) to~(\ref{eq:kappa_tr}), one obtains
\begin{equation}
\label{eq:kappa_tr1}
\kappa^{*}=\frac{1}{\Lambda^{\prime}_{\mathrm{P},1}(0) T}=-\frac{1}{\tilde{K}_{11}C_{11}+\tilde{K}_{21}C_{21}}.
\end{equation}

\subsection{\label{subsec:triv_fe_lim}Trivial Floquet exponent in the presence of $\Lambda_{\mathrm{D},n+1}(0)\rightarrow +0$}

Let us recall that our goal is put $\Lambda_{\mathrm{D},n+1}(0)$ close to zero and obtain the stabilization at the threshold control gain $\kappa^{*}\sim \mathcal{O}(\Lambda_{\mathrm{D},n+1}(0))$. Up to now, we did not use the property $\Lambda_{\mathrm{D},n+1}(0)\rightarrow +0$. As it is shown in Appendix~\ref{appsec:fm_and_prc}, $\mathbf{v}_{1,1:n} (t)$ is the PRC of the oscillator without the unstable degree of freedom $w(t)$; thus, $C_{11}$ does not depend on $\Lambda_{\mathrm{D},n+1}(0)$; i.e., $C_{11}\sim \mathcal{O}(1)$. Therefore, it turns out, that the only potentially successful choice of the control form is $\tilde{K}_{21} =\pm 1$.

Using~(\ref{eqapp:prc_n_p1_sol}) the $(n+1)$th component of the PRC $v_{1,n+1} (t)$ in the limit $\Lambda_{\mathrm{D},n+1}(0) \rightarrow +0$ reads
\begin{equation}
\label{eq:fm_np1}
\begin{aligned}
& v_{1,n+1} (t)= -\frac{\exp\left(-\lambda_{\mathrm{l}} t-\lambda_{\mathrm{n}}\int_0^t g(\bm{\xi}(t^{\prime}))\mathrm{d}t^{\prime}\right)}{\Lambda_{\mathrm{D},n+1}(0) T} \\
& \times \int\limits_0^T \mathbf{v}_{1,1:n}^T (t^{\prime})  \mathbf{p}(t^{\prime}) \exp\left(\lambda_{\mathrm{l}} t^{\prime}+\lambda_{\mathrm{n}}\int_0^{t^{\prime}} g(\bm{\xi}(t^{\prime \prime}))\mathrm{d}t^{\prime \prime}\right) \mathrm{d}t^{\prime}.
\end{aligned}
\end{equation}
Note that the last expression, strictly speaking, is not periodic; however, the order of non-periodicity $\left[ v_{1,n+1} (0)-v_{1,n+1} (T) \right]\sim \mathcal{O}(1)$ while the order of $v_{1,n+1}(t)$ itself is $\mathcal{O}(1/\Lambda_{\mathrm{D},n+1}(0))$. To finalize our findings, we substitute~(\ref{eq:fm_np1}) to~(\ref{eq:for_c_3}) and then to~(\ref{eq:kappa_tr1}) and obtain
\begin{equation}
\label{eq:kappa_tr2}
\kappa^{*} = \frac{\Lambda_{\mathrm{D},n+1}(0) T}{\tilde{K}_{21} I_1 I_2} ,
\end{equation}
where the integrals read
\begin{equation}
\label{eq:i1}
I_1=\left[ \int\limits_0^T \mathbf{q}^T(t) \dot{\bm{\xi}}(t) \exp\left(-\lambda_{\mathrm{l}} t-\lambda_{\mathrm{n}}\int_0^{t} g(\bm{\xi}(t^{\prime}))\mathrm{d}t^{\prime}\right) \mathrm{d}t \right],
\end{equation}
\begin{equation}
\label{eq:i2}
I_2=\left[ \int\limits_0^T \mathbf{v}_{1,1:n}^T (t)  \mathbf{p}(t) \exp\left(\lambda_{\mathrm{l}} t+\lambda_{\mathrm{n}}\int_0^{t} g(\bm{\xi}(t^{\prime}))\mathrm{d}t^{\prime}\right) \mathrm{d}t \right].
\end{equation}
As one can see, $\kappa^{*}\sim \mathcal{O}(\Lambda_{\mathrm{D},n+1}(0))$ if both integrals $I_1$ and $I_2$ have a non-zero zeroth-order term for an expansion with respect to $\Lambda_{\mathrm{D},n+1}(0)$. If $\lambda_{\mathrm{n}}=0$, then $\lambda_{\mathrm{l}}=\Lambda_{\mathrm{D},n+1}(0)$, and the integral $I_1$ up to the order $\mathcal{O}(1)$ gives
\begin{equation}
\label{eq:int}
\begin{aligned}
 I_1 &=  \int\limits_0^T \mathbf{q}^T(t) \dot{\bm{\xi}}(t) \mathrm{d}t = \ointctrclockwise\limits_{\bm{\xi}(t)} \left[ \nabla g(\mathbf{x}) \right]^T \cdot \mathrm{d}\mathbf{x} \\
& = g(\bm{\xi}(0))-g(\bm{\xi}(T))=0.
\end{aligned}
\end{equation}
Therefore, in order to have small threshold control gain $\kappa^{*}$, one should satisfy the constraints \hyperlink{cond1}{(i)} and \hyperlink{cond2}{(ii)}. In general, both integrals $I_1$ and $I_2$ are non-zero. However, for some cases, one of the integrals may be equal to zero for any parameter values $\lambda_{\mathrm{n}}$ and $\lambda_{\mathrm{l}}$ that give $\Lambda_{\mathrm{D},n+1}(0) \rightarrow +0$. As an example, the problematic situation may appear if $\int_0^{T} g(\bm{\xi}(t))\mathrm{d}t=0$. In Sec.~\ref{subsec:two_ls}, we analyze this case on the Landau-Stuart oscillator and show how to overcome this restriction.

The expression~($\ref{eq:kappa_tr2}$), derived by taking into account the conditions \hyperlink{cond1}{(i)} and \hyperlink{cond2}{(ii)}, shows that some branch of FE passes $\Lambda=0$, and it happens at the small control gain. However, we do not know whether this passing happens for the unstable branch of FE or for some other branch of FE. Therefore, up to now, we show that \hyperlink{cond1}{(i)} and \hyperlink{cond2}{(ii)} are necessary but not sufficient conditions in order to stabilize the limit cycle with small $\kappa$. In order to prove the sufficiency of both conditions, one should additionally analyze the unstable FE $\Lambda_{\mathrm{D},n+1}(\kappa)$ near $\kappa\rightarrow 0$.

\subsection{\label{subsec:uns_fe}Unstable Floquet exponent in the presence of $\Lambda_{\mathrm{D},n+1}(0)\rightarrow +0$}

The information on the derivative $\mathrm{d}\Lambda_{\mathrm{D},n+1}(\kappa)/\mathrm{d}\kappa$ near $\kappa=0$ can be extracted from the unstable FE of the PFC system $\Lambda_{\mathrm{P},n+1}(\Gamma)$ near $\Gamma=0$. According to~(\ref{eq:dp_relation}), for the unstable FE,
\begin{equation}
\label{eq:uns_dfc_pfc}
\Lambda_{\mathrm{D},n+1}(\kappa)=\Lambda_{\mathrm{P},n+1}\left( \kappa\left[ 1-\frac{1}{\exp(\Lambda_{\mathrm{D},n+1}(\kappa) T)} \right] \right).
\end{equation}
Moreover, unlike the case of the trivial FE, the points $\kappa\rightarrow0$ map to $\Gamma\rightarrow0$. Differentiation of (\ref{eq:uns_dfc_pfc}) with respect to $\kappa$ gives
\begin{equation}
\label{eq:uns_d}
\Lambda^{\prime}_{\mathrm{D},n+1}(0)= \left[1-\frac{1}{\exp(\Lambda_{\mathrm{D},n+1}(0)T)} \right] \Lambda^{\prime}_{\mathrm{P},n+1}(0),
\end{equation}
or in the limit $\Lambda_{\mathrm{D},n+1}(0)\rightarrow +0$, it simplifies to
\begin{equation}
\label{eq:uns_d1}
\Lambda^{\prime}_{\mathrm{D},n+1}(0)= T \Lambda_{\mathrm{D},n+1}(0) \Lambda^{\prime}_{\mathrm{P},n+1}(0).
\end{equation}
The last equation relates derivatives of FEs in the DFC and PFC systems. If the right hand side of~(\ref{eq:uns_d1}) is not small (of order of $\sim \mathcal{O}(1)$), then the stabilization may be achieved for a small control gain $\kappa$. Moreover, if the unstable FE's branch of the DFC system contains the trivial and unstable branches of the PFC system (as it is in the example of Fig.~\ref{fig:pfc_dfc} where two branches of the PFC system transform into one branch of the DFC system), then the threshold control gain~(\ref{eq:kappa_tr2}) is exactly the point where the unstable FE of the DFC system crosses zero. In such a situation, the derivative $\Lambda^{\prime}_{\mathrm{D},n+1}(0)$ can also be calculated as a finite difference (see Fig.~\ref{fig:pfc_dfc}~(b)),
\begin{equation}
\label{eq:uns_d2}
\Lambda^{\prime}_{\mathrm{D},n+1}(0)= - \frac{\Lambda_{\mathrm{D},n+1}(0)}{\kappa^*}.
\end{equation}
The main goal in this subsection is to show that the right hand side of Eq.~(\ref{eq:uns_d1}) is of the order of $\mathcal{O}(1)$ and to prove that (\ref{eq:uns_d2}) is valid if both conditions \hyperlink{cond1}{(i)} and \hyperlink{cond2}{(ii)} are satisfied. After these proofs, one can claim that the conditions \hyperlink{cond1}{(i)} and \hyperlink{cond2}{(ii)} will become necessary and sufficient to stabilize the limit cycle. 

Let us evaluate the derivative $\Lambda^{\prime}_{\mathrm{P},n+1}(0)$, which is on the right hand side of Eq.~(\ref{eq:uns_d1}). Similarly to the case of Eqs.~(\ref{eq:l_tr}) and~(\ref{eq:fm_tr}), we perform an expansion with respect to the control gain $\Gamma$
\begin{equation}
\label{eq:l_uns}
\Lambda_{\mathrm{P},n+1}(\Gamma)=\Lambda_{\mathrm{P},n+1}(0)+\Lambda^{\prime}_{\mathrm{P},n+1}(0) \Gamma+\mathcal{O}(\Gamma^2),
\end{equation}
\begin{equation}
\label{eq:fm_uns}
\mathbf{u}_{\mathrm{P},n+1}(t,\Gamma)=\mathbf{u}_{n+1}(t)+\mathbf{u}^{(1)}_{\mathrm{P},n+1}(t)\Gamma+\mathcal{O}(\Gamma^2).
\end{equation}
Next, by performing analogous steps as were done after Eq.~(\ref{eq:fm_tr}), we will end up with the linear form
\begin{equation}
\label{eq:anp1}
\begin{aligned}
&\Lambda^{\prime}_{\mathrm{P},n+1}(0) T = \\
&-\left( \tilde{K}_{11} B_{11}+\tilde{K}_{12} B_{12}+\tilde{K}_{21} B_{21}+\tilde{K}_{22} B_{22} \right),
\end{aligned}
\end{equation}
where the coefficients $B_{ij}$ read
\begin{subequations}
\label{eq:for_cp}
\begin{align}
B_{11} &= \int\limits_0^{T} \mathbf{v}_{n+1,1:n}^T (t)  \mathbf{p}(t) \mathbf{q}^T(t) \mathbf{u}_{n+1,1:n}(t) \mathrm{d}t, \label{eq:for_cp_1} \\
B_{12} &= \int\limits_0^{T} \mathbf{v}_{n+1,1:n}^T (t)  \mathbf{p}(t) u_{n+1,n+1}(t) \mathrm{d}t, \label{eq:for_cp_2} \\
B_{21} &= \int\limits_0^{T} v_{n+1,n+1}(t) \mathbf{q}^T(t) \mathbf{u}_{n+1,1:n}(t)  \mathrm{d}t, \label{eq:for_cp_3} \\
B_{22} &= \int\limits_0^{T} v_{n+1,n+1}(t) u_{n+1,n+1}(t)  \mathrm{d}t. \label{eq:for_cp_4}
\end{align}
\end{subequations}
As it is shown in Appendix~\ref{appsec:np1_lfm}, $\mathbf{v}_{n+1,1:n} (t)=0$ and $v_{n+1,n+1}(t) u_{n+1,n+1}(t)=1$. Therefore, $B_{11}=B_{12}=0$ and $B_{22}=T$. Again, one can conclude that the only possibility to have $\Lambda^{\prime}_{\mathrm{P},n+1}(0) \sim \mathcal{O}(\Lambda^{-1}_{\mathrm{D},n+1}(0))$ and as a consequence to have the right hand side of (\ref{eq:uns_d1}) order of $\mathcal{O}(1)$ is to set non-zero coefficient $\tilde{K}_{21}$.

Let us estimate the coefficient $B_{21}$ in the limit ${\Lambda_{\mathrm{D},n+1}(0)\rightarrow +0}$. As it is shown in Appendix~\ref{appsec:np1_rfm} for the given limit,  the right Floquet mode $\mathbf{u}_{n+1,1:n}(t)$ is defined by Eq.~(\ref{eqapp:nhoms3}), while $v_{n+1,n+1}(t)$ according to Eq.~(\ref{eqapp:lfm}) reads as $v_{n+1,n+1}(t)=\exp\left[- \lambda_{\mathrm{l}}t - \lambda_{\mathrm{n}} \int\limits_0^t g \left(\bm{\xi}(t^{\prime}) \right) \mathrm{d}t^{\prime} \right]$. Thus, $B_{21}=I_1 I_2/\left[ \Lambda_{\mathrm{D},n+1}(0) T \right] $, and by substituting Eq.~(\ref{eq:anp1}) into Eq.~(\ref{eq:uns_d1}), one can see that $\Lambda^{\prime}_{\mathrm{D},n+1}(0)=-\tilde{K}_{21} I_1 I_2/T$ is of the order of $\mathcal{O}(1)$ if both integrals $I_1$ and $I_2$ are non-zero. Additionally, using the expression~(\ref{eq:kappa_tr2}), one can conclude that Eq.~(\ref{eq:uns_d2}) holds.

\subsection{\label{subsec:summ_lim}Summary and limitations}

Let us summarize Sec.~\ref{sec:udfc_oscillator}. The stable limit cycle oscillator possessing single input and single output can be destabilized by the unstable degree of freedom and then stabilized by the DFC force. The equations for the oscillator under control read
\begin{subequations}
\label{eq:one_osc_udfcf}
\begin{align}
\dot{\mathbf{x}} &= \mathbf{f} \left(\mathbf{x}, r \right), \label{eq:one_osc_udfcf_1} \\
\dot{w} &= \left( \lambda_{\mathrm{l}}+\lambda_{\mathrm{n}} s(t) \right) w + \kappa  \tilde{K}_{21} \left[ s(t-\tau)-s(t) \right] , \label{eq:one_osc_udfcf_2} \\
s(t) &= g \left(\mathbf{x}(t) \right), \label{eq:one_osc_udfcf_3} \\
r(t)&= w(t) .\label{eq:one_osc_udfcf_4}
\end{align}
\end{subequations}
Here, the time delay $\tau=T$ and $\tilde{K}_{21}=\pm 1$ where, for the convenience, the appropriate sign can be chosen in such a way that the threshold control gain $\kappa^{*}$ is always positive. The linear $ \lambda_{\mathrm{l}}$ and non-linear $\lambda_{\mathrm{n}}$ contributions to the unstable degree of freedom can be adjusted in such a way that the induced unstable FE $\Lambda_{n+1}$ will be close to zero (a weak instability). In order to have stable system~(\ref{eq:one_osc_udfcf}), one should have non-zero integrals (\ref{eq:i1}) and (\ref{eq:i2}), and as a consequence, $\lambda_{\mathrm{n}}$ should be non-zero. The stability appears when the control gain $\kappa$ is slightly above to the threshold value $\kappa^{*}$ given by~(\ref{eq:kappa_tr2}). Note that a further increase of $\kappa$ does not guarantee the stability of the limit cycle. As an example, see Fig.~\ref{fig:FHN_mu}, where for $\kappa>\kappa^{*}\approx 2.3 \times 10^{-4}$, the limit cycle becomes stable, but for $\kappa> 12.5 \times 10^{-4}$, the limit cycle again loses its stability.

The main consequence of the control in the system~(\ref{eq:one_osc_udfcf}) is that for the case of weakly coupled oscillators, the phase $\psi(t)$ of the controlled oscillator behaves in such a way like the coupling constant $\varepsilon$ will have an opposite sign, $\varepsilon_{\mathrm{eff}}=\varepsilon \alpha\left(\kappa \tilde{K}_{21} C_{21}\right) $ (cf. Eq.~(\ref{eq:eps_eff})) with negative $\alpha$ and the constant $C_{21}=-I_1 I_2/(\Lambda_{n+1}T)$. One of the prominent examples is when $\kappa=2\kappa^{*}$. Then, the factor $\alpha=-1$ and as a consequence, $\varepsilon_{\mathrm{eff}}=-\varepsilon$. If the delayed term $s(t-\tau)$ in Eq.~(\ref{eq:one_osc_udfcf_2}) has time delay $\tau$ close but not equal to the natural oscillator's period $T$, then, according to Eq.~(\ref{eq:omega_eff}), the natural frequency of the phase model changes to the effective frequency $\omega_{\mathrm{eff}}=\omega+(T-\tau)\Omega\left[1-\alpha\left(\kappa \tilde{K}_{21}C_{21}\right) \right]/T$.

The control method~(\ref{eq:one_osc_udfcf}) has two hidden limitations. The first limitation is related to the two small parameters $\varepsilon$ and $\Lambda_{n+1}$ and an interference between them. The phase reduction theory is derived in the limit ${\varepsilon \rightarrow 0}$, where $\varepsilon$ is an inter-oscillatory coupling constant (see Eq.~(\ref{eq:main})), while the results of this section are derived for $\Lambda_{n+1}\rightarrow +0$. In order to have a valid phase dynamics governed by Eq.~(\ref{eq:phase_cont}), an interference between these two small values should be avoided. In particular, one should set the unstable FE $\Lambda_{n+1}$ larger than the inter-oscillatory coupling constant $\varepsilon$. Otherwise, the phase reduction will no longer work. As it will be shown below in Sec.~\ref{sec:two_osc_in_anti}, in order to avoid such interference, we chose the unstable FE $\Lambda_{n+1}$ about 5 times higher than $\varepsilon$.

The second limitation is related to an additional periodic orbit (not equal to the analyzed limit cycle $(\mathbf{x}^T (t),w(t))=(\bm{\xi}^T (t),0)$) induced due to the unstable degree of freedom and the DFC force. The profile of the additional periodic orbit is derived in Appendix~\ref{appsec:add_per_orbit}. One can think that it is convenient to choose the control gain $\kappa$ just slightly above the threshold gain $\kappa^{*}$ since the limit cycle is certainly stable and the effective inter-oscillatory coupling constant $\varepsilon_{\mathrm{eff}}$ is highly negative. However the additional periodic orbit restricts the choice of $\kappa$ being close to $\kappa^{*}$. By increasing $\kappa$ from zero, the control force does not change the profile and period of the limit cycle $(\bm{\xi}^T (t),0)$, while the additional periodic orbit changes its profile and period. At the point $\kappa=\kappa^{*}$, two orbits collide (transcritical bifurcation of periodic orbits) and interchange their stability. The additional orbit is stable before $\kappa$ achieves $\kappa^{*}$ and is unstable after $\kappa$ increases further. Thus, for $\kappa\rightarrow\kappa^{*}+0$ in the phase space near the stable limit cycle $(\bm{\xi}^T (t),0)$, there is unstable orbit that repels solutions; therefore, the basin of attraction of the limit cycle $(\bm{\xi}^T (t),0)$ becomes very narrow. Once the oscillator~(\ref{eq:one_osc_udfcf}) is coupled to other oscillators, it receives perturbations, which can push out the state from the basin of attraction. Such a scenario leads to uncontrollable growth of the variable $w(t)$, indicating that the controller is no longer relevant. In order to avoid such a scenario, one should set the control gain $\kappa$ far away from the threshold gain $\kappa^{*}$. Additionally, it potentially prevents from being $\Lambda_{\mathrm{D},n+1}(\kappa)$ close to zero.

\section{\label{sec:two_osc_in_anti}Switching between in-phase and anti-phase synchronization for two weakly coupled oscillators}

The near-identical two limit cycle oscillators coupled attractively (repulsively) with a large enough coupling constant demonstrate in-phase (anti-phase) synchronization. By applying the unstable DFC to both oscillators, one can reverse the sign of the coupling constant and switch between the in-phase and anti-phase synchronous regimes. In this section, we present a numerical demonstration of this switching on two different oscillator models, namely, FitzHugh-Nagumo and Landau-Stuart.

\subsection{\label{subsec:two_fhn}Numerical demonstration on two FitzHugh-Nagumo neurons}

In the first demonstration, we will use FitzHugh-Nagumo oscillators. The dynamics of the $i$th oscillator is described by the following vector field:
\begin{equation}
\mathbf{f}^{(i)}\left(\mathbf{x},r \right)=
\left[
\begin{array}{c}
x_{1} - x_{1}^3/3 -x_{2} + 0.5   \\
\epsilon^{(i)} \left[  (1+r)x_{1}+0.7 - 0.8 x_{2} \right] 
\end{array}
\right].
\label{fhn_f}
\end{equation}
Here, $x_{j}$ denotes the $j$th component of the vector $\mathbf{x}$. The oscillators differ by parameter $\epsilon^{(i)}$, which also defines the natural frequency. The coupling function $\mathbf{G}^{(ij)}\equiv \mathbf{G}$ is set through the first dynamical variables by a non-trivial form
\begin{equation} 
\mathbf{G}(\mathbf{y},\mathbf{x})= 
\left[
\begin{array}{c}
y_{1} /\left( 2+ y_{2} \right) -  x_{1} / \left ( 2+ x_{2} \right) \\
0 
\end{array}
\right].
\label{fhn_G}
\end{equation}
We assume that the measured scalar signal has the form
\begin{equation} 
s(t)=g(\mathbf{x}(t))= x^2_{1}(t)+x_{2}(t)
\label{fhn_s}
\end{equation}
consisting both variables of the system. Such unusual forms are chosen in order to demonstrate the universality of the proposed algorithm. 

\begin{figure}[h!]
	\centering\includegraphics[width=0.95\columnwidth]{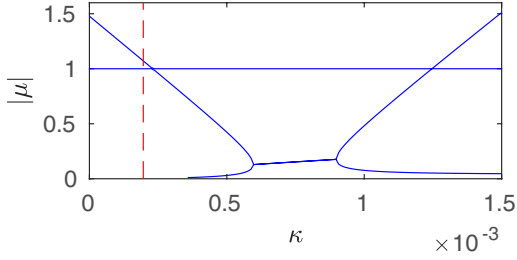}
	\caption{\label{fig:FHN_mu} The dependence of the FMs on the control gain in the case of DFC for the FitzHugh-Nagumo oscillator model. The control form $\tilde{\mathbf{K}}$ has non-zero element $\tilde{K}_{21}=-1$, and the linear and non-linear contributions to the dynamics of the unstable degree of freedom are $\lambda_{\mathrm{l}}=0.595$ and $\lambda_{\mathrm{n}}=-0.25$. The red dashed line calculated from Eq.~(\ref{eq:kappa_tr2}) is close to the numerically obtained critical control gain $\kappa^* \approx 2.3 \times 10^{-4}$. Note that the unstable branch of the FM can be reproduced from the unstable branch of the FE in Fig.~\ref{fig:pfc_dfc}(b).}
\end{figure}

Firstly, we investigate the effect of the DFC force for the single oscillator with the additional unstable degree of freedom. For that purpose in Fig.~\ref{fig:FHN_mu}, we plot dependence of the absolute value of FMs on the control gain $\kappa$. Here, we set the intrinsic system parameter $\epsilon = 0.08$ and parameters of the control $\lambda_{\mathrm{l}} = 0.595$, $\lambda_{\mathrm{n}}= -0.25$, $\tilde{K}_{21} = -1$. At $\kappa=0$, the unstable FM is $|\mu_3| \approx 1.48$ (FE is $\Lambda_3 \approx 0.01$ and oscillator period $T \approx 39.4744$). By increasing $\kappa$, $|\mu_3|$ decreases, and at $\kappa=\kappa^* \approx 2.3\times 10^{-4}$, it becomes equal to one. The oscillator remains stable for interval $\kappa \in [2.3 ,12.5] \times 10^{-4}$.

\begin{figure}[h!]
	\centering\includegraphics[width=0.95\columnwidth]{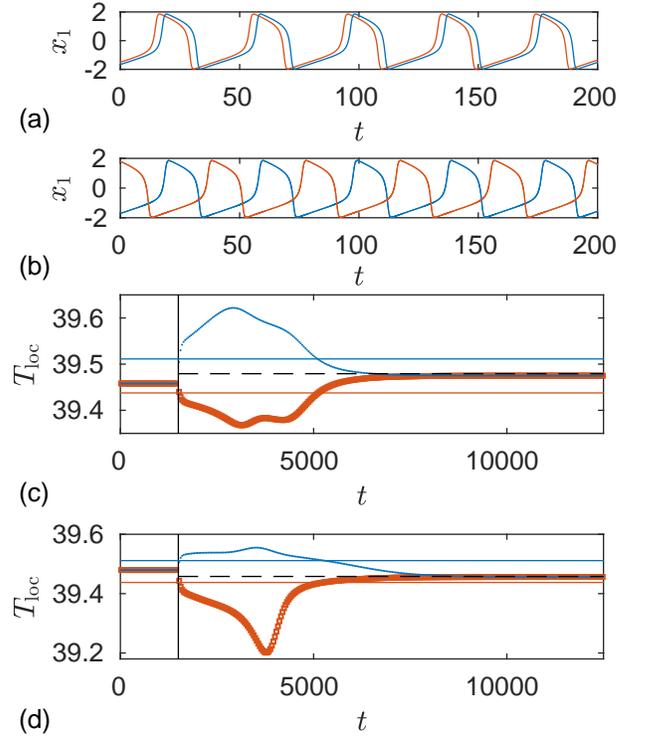}
	\caption{\label{fig:FHN} Effect of DFC for the two coupled FitzHugh-Nagumo neurons. In (a)-(c) the coupling strength is $\varepsilon = 2 \times 10^{-3}$ and in (d)  $\varepsilon = -2 \times 10^{-3}$. (a) Shows dynamics of the first dynamical variables of oscillators in the control-free case; (b) the same as in (a), but the system is under control; and (c) and (d) ``local'' periods for the control-free case (before a vertical black line) and under control (after a vertical black line). The horizontal solid lines represent intrinsic periods of oscillators, $T^{(1)}$ and $T^{(2)}$, while the horizontal dashed line represents a synchronization period obtained from the phase model. Parameters:  $\epsilon^{(i)} =0.08+\delta\epsilon_i \times 10^{-4}$ where $\delta\bm{\epsilon}=[1,-1]^T$, $\lambda_{\mathrm{l}} = 0.595$, $\lambda_{\mathrm{n}}= -0.25$, $\tilde{K}_{21} = -1$, $\kappa = 4.6 \times 10^{-4}$, $\tau^{(i)}=T^{(i)}$. }
\end{figure}

Secondly, we couple two neurons \eqref{fhn_f} with $\epsilon^{(i)} =0.08+[1, -1] \times 10^{-4}$. The natural periods of oscillators are $T^{(1)} \approx 39.4376$ and  $T^{(2)} \approx 39.5115$.  When coupling strength is set to $\varepsilon = 2 \times 10^{-3}$, neurons are nearly in-phase synchronized; see dynamics of the first variables of neurons in Fig.~\ref{fig:FHN}(a). When $\varepsilon = -2 \times 10^{-3}$ neurons are in an anti-phase synchronization regime.  We apply control algorithm \eqref{eq:one_osc_udfcf} for both neurons, with $\lambda_{\mathrm{l}}$ and $\lambda_{\mathrm{n}}$ values used in Fig.~\ref{fig:FHN_mu}, time delay $\tau^{(i)}=T^{(i)}$ and $\kappa= 2\kappa^{*} = 4.6 \times 10^{-4}$. The choice of $\kappa$ determines that $\varepsilon_{\mathrm{eff}} = -\varepsilon$. In Fig.~\ref{fig:FHN}(b), the dynamics of oscillators is plotted after transitional processes when $\varepsilon = 2 \times 10^{-3}$ and the system is under control. We see that neurons are in nearly anti-phase synchronization. We calculate the time distances between two neighboring maximums of the first dynamical variables and call this quantity ``local'' period $T^{(i)}_{\mathrm{loc}}$. For example, in Fig.~\ref{fig:FHN}(a), the blue curve achieves maxima at $t_1 \approx 19$ and $t_2 \approx 59$. Thus, at the time moment $t_1$, the ``local'' period $T^{(i)}_{\mathrm{loc}}(t_1)=t_2-t_1$. The ``local'' periods are plotted as discrete symbols, but on a broad time interval, they form a continuous curve. When the system is in the frequency-locking regime, both $T^{(i)}_{\mathrm{loc}}$ coincide and form a horizontal line. In Fig.~\ref{fig:FHN}(c), we plot ``local'' periods for a control-free system when $0<t<10^4$ and with control being turned on at $t>10^4$. We see that after transitional processes, both $T^{(i)}_{\mathrm{loc}}$ become equal and coincide with the synchronization period obtained from the phase model (dashed line). The same behavior is seen in Fig.~\ref{fig:FHN}(d), where the control-free system for $\varepsilon = -2 \times 10^{-3}$ was in an anti-phase synchronization regime, and after the control is turned on at $t=10^4$, the system becomes in-phase synchronized.

\subsection{\label{subsec:two_ls}Numerical demonstration on two Landau-Stuart oscillators}

The dynamics of the $i$th Landau-Stuart oscillator is described by the following vector field:
\begin{equation}
\mathbf{f}^{(i)}\left(\mathbf{x}, r \right)=
\left[
\begin{array}{c}
x_{1} \left(1-x_{1}^2-x_{2}^2 \right) - \Omega^{(i)} x_{2} \\
x_{2} \left(1-x_{1}^2-x_{2}^2 \right) + \Omega^{(i)} x_{1} + r
\end{array}
\right].
\label{ls_f}
\end{equation}
Here, $x_{j}$ denotes the $j$th component of the vector $\mathbf{x}$. The parameter $\Omega^{(i)}$ is a natural frequency of the $i$th oscillator. The coupling is realized through the first dynamical variables,
\begin{equation} 
\mathbf{G}(\mathbf{y},\mathbf{x})= 
\left[
\begin{array}{c}
y_{1} -x_{1} \\
0 
\end{array}
\right]
.\label{ls_G}
\end{equation}
We assume that only the first component of the oscillator is  available for the measurement,
\begin{equation} 
s(t)=g(\mathbf{x}(t))= x_{1}(t).
\label{ls_s}
\end{equation}

Due to simplicity of the Landau-Stuart model, the solution of a single  oscillator with frequency $\Omega$ can be written as ${\bm{\xi} (t)=[\cos(\Omega t), \sin(\Omega t)]^T}$ and the PRC is ${\mathbf{v}_{1,1:2}(t) = \Omega^{-1}[-\sin(\Omega t), \cos(\Omega t)]^T}$. Let us set $\Omega=1$. According to~(\ref{ls_s}), we are measuring the first dynamical variable; therefore, the unstable FE $\Lambda_{\mathrm{D},3}(0) = \lambda_{\mathrm{l}}$ does not depend on $\lambda_{\mathrm{n}}$ since the integral $\int_0^T g \left(\bm{\xi}(t) \right) \mathrm{d}t =0$. This means that if we want to achieve $\Lambda_{\mathrm{D},3}(0) \to +0$, we have also to take $  \lambda_{\mathrm{l}} \to +0$. As a consequence  the integral $I_2$ of Eq.~\eqref{eq:i2} becomes $\int_{0}^{2\pi} \cos(t) \exp(\lambda_{\mathrm{n}} \sin(t)) \mathrm{d}t=0$ for any value of $\lambda_{\mathrm{n}}$. In other words the critical coupling $\kappa^*$ becomes infinite. Note that this problem is unusual and appears because the Landau-Stuart oscillator possesses high order symmetries, and the single-input variable $r$ affects the second dynamical variable. For example, if $r$ will be added to the first component of Eq.~(\ref{ls_f}), the problem will no longer appear.

The problem can be overcome by modifying Eq.~\eqref{eq:one_osc_udfcf_2} by the following scheme:
\begin{equation} 
\dot{w} = \left( \lambda_{\mathrm{l}}+\lambda_{\mathrm{n}} s_{\mathrm{tr}}(t) \right) w + \kappa  \tilde{K}_{21} \left[ s(t-T)-s(t) \right]. 
\end{equation}
Here, $s_{\mathrm{tr}}(t)$ is transformed signal $s(t)$. Nonlinear transformation may prevent annihilation of the integral $I_2$. In our case, the signal $s(t)$ varies in an interval $[-1;1]$. The simplest nonlinear transformation that also varies in this interval is the following quadratic function:
\begin{equation}
s_{\mathrm{tr}}(t)=0.5 s^2(t)+s(t)-0.5.
\end{equation} 

\begin{figure}[h!]
	\centering\includegraphics[width=0.95\columnwidth]{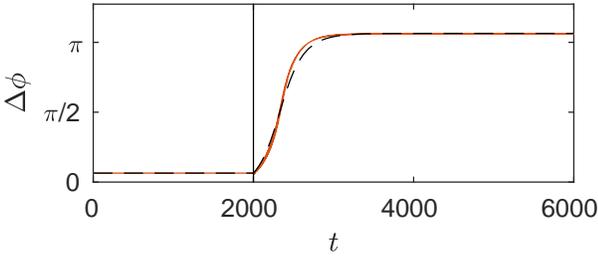}
	\caption{\label{fig:LS} Effect of the DFC for the coupled Landau-Stuart oscillators. The solid line is calculated directly from the models~Eq.(\ref{ls_f}), while the dashed line is obtained from the phase models. Parameters: $\Omega^{(1)} = 1$, $\Omega^{(2)}=1.001$, $\varepsilon = 5\times 10^{-3}$, $\lambda_{\mathrm{l}} = -0.3$, $\lambda_{\mathrm{n}} = -1.6$ ($\Lambda_3 = 0.1$, $\mu_3=1.87$), $\tilde {K}_{21}=-1$, $\kappa = 0.14$, $\tau^{(i)}=T^{(i)}$. The control is turned on at $t=2000$.}
\end{figure}

To demonstrate the effect of the modified version of the algorithm  on the Landau-Stuart model, we estimate oscillators phase $\phi(t)$ as an argument of a complex number $x_1(t) +\mathrm{i} x_2(t)=\rho(t)\mathrm{e}^{\mathrm{i}\phi(t)}$. In Fig.~\ref{fig:LS}, we show how the phase difference between two oscillators evolves in a control-free case  $t<2000$ and under control $t>2000$. We see that without control, the system was near the in-phase synchronization regime, and after the control is turned on, the oscillators reach a near anti-phase synchronization regime. The time delay $\tau^{(i)}=T^{(i)}$ and the control gain $\kappa$ were set in such a way that the effective coupling strength $\varepsilon_{\mathrm{eff}} = -\varepsilon$.

\section{\label{sec:non-inv}Non-invasive controller to stabilize an unstable phase difference of two heterogeneous oscillators}

In Sec.~\ref{sec:two_osc_in_anti}, we show the possibility of the unstable delayed feedback controller to achieve the anti-phase or in-phase synchronization regime when the control force is invasive. To be more precise, the delay times $\tau^{(i)}$ are chosen to be equal to the natural periods $T^{(i)}$ of the oscillators, while the achieved synchronous regime has its own synchronization period, which, in general, differs from $T^{(i)}$ by $\varepsilon$ order quantity. It means that $\left[ s^{(i)}(t-T^{(i)})-s^{(i)}(t) \right]\sim \mathcal{O}(\varepsilon)$, and therefore, the control force $r(t)=w(t)$ is $\varepsilon$-invasive. If two oscillators are coupled strongly enough such that in the control-free regime they are synchronized (frequency locking) and have some phase difference, then, according to phase reduction, there should be another phase difference, for which two oscillators are also synchronized. However, this phase difference is unstable. The goal of this section is to show that the unstable DFC is able to stabilize the unstable phase difference when this phase difference is not known a priori. Moreover, it can be done by the control force for which the $\mathcal{O}(\varepsilon)$ order term vanishes. Therefore, we refer to such controller as a non-invasive. Although $\mathcal{O}(\varepsilon^j)$ order terms with $j\geq 2$ do not necessary vanish, one should take into account that all consequences of the conventional phase reduction are valid only up to $\mathcal{O}(\varepsilon)$ order, and therefore, the existence of this unstable phase difference can be proved only up to $\mathcal{O}(\varepsilon)$.

Let us start with two weakly coupled control-free heterogeneous limit cycle oscillators,
\begin{subequations}
\label{eq:two_contr_free}
\begin{align}
\dot{\mathbf{x}}^{(1)} &= \mathbf{f}^{(1)} \left(\mathbf{x}^{(1)} \right)+\varepsilon \mathbf{G}^{(12)}\left( \mathbf{x}^{(2)} , \mathbf{x}^{(1)}  \right), \label{eq:two_contr_free_1} \\
\dot{\mathbf{x}}^{(2)} &= \mathbf{f}^{(2)} \left(\mathbf{x}^{(2)} \right)+\varepsilon \mathbf{G}^{(21)}\left( \mathbf{x}^{(1)} , \mathbf{x}^{(2)}  \right). \label{eq:two_contr_free_2}
\end{align}
\end{subequations}
Here, both vector fields $\mathbf{f}^{(1)}$ and $\mathbf{f}^{(2)}$, in general, can be dissimilar and even can have different dimensions. The only requirement is that the difference of the natural periods $T^{(1)}-T^{(2)}$ should be of the order of $\varepsilon$. By applying phase reduction, in the reference frame rotating with the ``central'' frequency $\Omega=2\pi/T$, the phase model reads
\begin{subequations}
\label{eq:ph_contr_free}
\begin{align}
\dot{\psi}^{(1)} &= \omega^{(1)}+ \varepsilon H^{(12)}\left( \psi^{(2)}-\psi^{(1)}  \right), \label{eq:ph_contr_free_1} \\
\dot{\psi}^{(2)} &= \omega^{(2)}+ \varepsilon H^{(21)}\left( \psi^{(1)}-\psi^{(2)}  \right). \label{eq:ph_contr_free_2}
\end{align}
\end{subequations}
Here, $H^{(ij)}$ is defined by Eq.~(\ref{eq:H_ij}). We assume that $\varepsilon$ is large enough such that the phase model~(\ref{eq:ph_contr_free}) possesses a frequency-locking solution. Therefore, the phase difference $\Delta \psi=\psi^{(2)}-\psi^{(1)}$ is governed by
\begin{equation}
\label{eq:del_phi}
\frac{\mathrm{d}}{\mathrm{d}t}\Delta\psi=\omega^{(2)}-\omega^{(1)}+\varepsilon h\left(\Delta\psi\right),
\end{equation} 
where
\begin{equation}
\label{eq:h_def}
h\left(\Delta\psi\right)=H^{(21)}\left(-\Delta\psi\right)-H^{(12)}\left(\Delta\psi\right),
\end{equation} 
should have at least one stable fixed point $\Delta\psi=\Delta\psi^{*}_s$ and one unstable fixed point $\Delta\psi=\Delta\psi^{*}_u$. Both points give the vanishing right hand side of Eq.~(\ref{eq:del_phi}),
\begin{equation}
\label{eq:dpsi_su}
h\left(\Delta\psi^{*}_{\left\lbrace s,u \right\rbrace} \right)=\frac{\omega^{(1)}-\omega^{(2)}}{\varepsilon},
\end{equation} 
while derivatives of the right hand side of Eq.~(\ref{eq:del_phi}) with respect to $\Delta\psi$ satisfy
\begin{subequations}
\label{eq:der_del_phi}
\begin{align}
\varepsilon h^{\prime}\left( \Delta\psi^{*}_s \right)<0, \label{eq:der_del_phi_1} \\
\varepsilon h^{\prime}\left( \Delta\psi^{*}_u \right)>0. \label{eq:der_del_phi_2}
\end{align}
\end{subequations}
If the system~(\ref{eq:ph_contr_free}) is in the fixed point $\Delta\psi^{*}_u$, both oscillators oscillate with the synchronization period $T_u$. It can be obtained by substituting $\Delta\psi^{*}_u$ into Eq~(\ref{eq:ph_contr_free_1}) or (\ref{eq:ph_contr_free_2}),
\begin{equation}
\label{eq:t_u}
\begin{aligned}
T_u &= T^{(1)}\left(1-\frac{\varepsilon}{\Omega} H^{(12)}\left(\Delta\psi^{*}_u\right) \right) \\
&=T^{(2)}\left(1-\frac{\varepsilon}{\Omega} H^{(21)}\left(-\Delta\psi^{*}_u\right) \right).
\end{aligned}
\end{equation} 
Our aim is to use unstable DFC for both oscillators with the time delays $\tau^{(1)}=\tau^{(2)}=T_u$ such that $\Delta\psi^{*}_u$ becomes a stable fixed point. Since at that point the output of the oscillator $s^{(i)}\left(t+T_u\right)=s^{(i)}\left(t\right)+\mathcal{O}(\varepsilon^2)$, the DFC force and the unstable degree of freedom $w$ vanish at the order $\mathcal{O}(\varepsilon)$.

Now, let us analyze the same two coupled heterogeneous oscillators~(\ref{eq:two_contr_free}), but under the unstable DFC,
\begin{subequations}
\label{eq:two_contr}
\begin{align}
\dot{\mathbf{x}}^{(1)} &= \mathbf{f}^{(1)} \left(\mathbf{x}^{(1)}, r^{(1)} \right)+\varepsilon \mathbf{G}^{(12)}\left( \mathbf{x}^{(2)} , \mathbf{x}^{(1)}  \right), \label{eq:two_contr_1} \\
\dot{\mathbf{x}}^{(2)} &= \mathbf{f}^{(2)} \left(\mathbf{x}^{(2)}, r^{(2)}\right)+\varepsilon \mathbf{G}^{(21)}\left( \mathbf{x}^{(1)} , \mathbf{x}^{(2)}  \right), \label{eq:two_contr_2}
\end{align}
\end{subequations}
where the single inputs $r^{(1)}$ and $r^{(2)}$ are constructed with respect to described algorithm~(\ref{eq:one_osc_udfcf}). We assume that both time delays $\tau^{(1)}=\tau^{(2)}=T_u$ and the control gains $\kappa^{(1)}$ and $\kappa^{(2)}$ are chosen in such a way that $\alpha^{(1)}$ and $\alpha^{(2)}$ are negative. It is convenient to set $\alpha^{(1)}=\alpha^{(2)}$; however, in a typical experimental setup due to the heterogeneity, it can be difficult to achieve. Therefore, we additionally  assume that $\alpha^{(1)}=\alpha$ and $\alpha^{(2)}=\alpha+\Delta\alpha$, where $\Delta\alpha$ is small (the smallness will be defined later). By performing the same steps, as in derivation of Eqs.~(\ref{eq:ph_contr_free}), one gets
\begin{subequations}
\label{eq:ph_contr}
\begin{align}
\dot{\psi}^{(1)} =& \omega^{(1)}+\Omega\frac{T_u-T^{(1)}}{T}\left[\alpha-1 \right] \nonumber \\ &+ \varepsilon \alpha H^{(12)}\left( \psi^{(2)}-\psi^{(1)}  \right), \label{eq:ph_contr_1} \\
\dot{\psi}^{(2)} =& \omega^{(2)}+\Omega\frac{T_u-T^{(2)}}{T}\left[\alpha+\Delta\alpha-1 \right] \nonumber \\
&+ \varepsilon \left[ \alpha+\Delta\alpha \right] H^{(21)}\left( \psi^{(1)}-\psi^{(2)}  \right). \label{eq:ph_contr_2}
\end{align}
\end{subequations}
By using Eq.~(\ref{eq:t_u}), given equations become
\begin{subequations}
\label{eq:ph_contr_v1}
\begin{align}
\dot{\psi}^{(1)} =& \omega^{(1)}+\varepsilon  H^{(12)}\left( \Delta\psi_u^{*}  \right)\left[1-\alpha \right] \nonumber \\ &+ \varepsilon \alpha H^{(12)}\left( \Delta\psi  \right), \label{eq:ph_contr_1_v1} \\
\dot{\psi}^{(2)} =& \omega^{(2)}+\varepsilon H^{(21)}\left( -\Delta\psi_u^{*}  \right)\left[1-\alpha-\Delta\alpha \right] \nonumber \\
&+ \varepsilon \left[ \alpha+\Delta\alpha \right] H^{(21)}\left( -\Delta\psi  \right). \label{eq:ph_contr_2_v1}
\end{align}
\end{subequations}
By using Eqs.~(\ref{eq:h_def}) and (\ref{eq:dpsi_su}), one can derive the dynamics for the phase difference $\Delta\psi$,
\begin{equation}
\label{eq:del_phi_cont}
\begin{aligned}
\frac{\mathrm{d}}{\mathrm{d}t}\Delta\psi =& \alpha\left[\omega^{(2)}-\omega^{(1)}+\varepsilon h\left(\Delta\psi\right)\right] \\
&+\varepsilon \Delta\alpha \left\lbrace  H^{(21)}\left(-\Delta\psi\right)- H^{(21)}\left(-\Delta\psi^{*}_u\right) \right\rbrace.
\end{aligned}
\end{equation} 
From Eq.~(\ref{eq:del_phi_cont}), one can see that it possesses the fixed point at $\Delta\psi=\Delta\psi^{*}_u$, and the stability of such a point is fulfilled if the inequality
\begin{equation}
\label{eq:der_del_phi_cont}
\alpha\varepsilon h^{\prime}\left( \Delta\psi^{*}_u \right)-\varepsilon \Delta\alpha \left. \frac{\mathrm{d} H^{(21)}(\chi)}{\mathrm{d}\chi}\right|_{\chi=-\Delta\psi^{*}_u}<0
\end{equation}
holds. The first term is always negative due to the inequality~(\ref{eq:der_del_phi_2}). If $\Delta\alpha$ is small enough such that the second term does not damage the inequality~(\ref{eq:der_del_phi_cont}), we end up with stabilization of the fixed point $\Delta\psi^{*}_u$. Interestingly, for the case of $\Delta\alpha=0$, Eq.~(\ref{eq:del_phi_cont}) resembles the control-free evolution of the phase difference governed by Eq.~(\ref{eq:del_phi}) with reversed time flow $t\rightarrow -t$ and that is the reason of flipping the stability of both fixed points.

We can conclude that the provided unstable DFC scheme is able to stabilize the unstable phase difference $\Delta\psi^{*}_u$, which exists in the phase model~(\ref{eq:del_phi}), and such stabilization is achieved by the non-invasive control force. In Appendix~\ref{appsec:non_inv}, we show that the control force expanded in the powers of $\varepsilon$ demonstrates the non-invasiveness up to $\mathcal{O}\left(\varepsilon^2\right)$. Our algorithm requires some sophisticated setup of the time delays. In particular, the time delays should be set $\tau^{(1)}=\tau^{(2)}=T_u$, while $T_u$ together with $\Delta\psi^{*}_u$ and $h(\cdot)$ is assumed to be unknown. However, the fact of the non-invasiveness of the control force at the point $\tau^{(1,2)}=T_u$ and invasiveness at the point $\tau^{(1,2)}\neq T_u$ can be exploited to set up the time delays. For example, in Sec.~\ref{subsec:two_het}, we employ the adaptive version of the DFC scheme where the time delays are slowly changed in time in such a way that the control force is minimized.

For the typical situation, when two oscillators are near-identical with identical couplings $\mathbf{G}^{(12)}\left( \mathbf{x},\mathbf{y} \right)=\mathbf{G}^{(21)}\left( \mathbf{x},\mathbf{y} \right)=\mathbf{G}\left( \mathbf{x},\mathbf{y} \right)$ and small enough dissimilarity of the frequencies $\left(\omega^{(2)}-\omega^{(1)}\right)/\varepsilon\rightarrow 0$, the stable and unstable fixed points satisfy $\Delta\psi^{*}_s-\Delta\psi^{*}_u\approx\pi$. A similar situation is analyzed in so-called equivariant DFC~\cite{Fiedler2010,Schneider2013,Schneider2016} where $N$ coupled identical units can possess an unstable spatiotemporal synchronization pattern, and such a pattern can be stabilized by the non-invasive control force of the form $s^{(n)}(t-T/N)-s^{(n+1)}(t)$. For the case of $N=2$, the vanishing control force $s^{(1)}(t-T/2)-s^{(2)}(t)=0$ means that the anti-phase solution is stabilized; therefore, $\Delta\psi^{*}_s-\Delta\psi^{*}_u=\pi$. The equivariant DFC and the provided unstable DFC pursue similar goals; therefore, it is interesting to discuss the differences between both algorithms. The case of equivariant DFC works for the symmetrical vector fields $\mathbf{f}^{(i)}$, which should be identical for all units, i.e., $\mathbf{f}^{(i)}=\mathbf{f}^{(j)}$, and imposes symmetric restrictions on the coupling function $\mathbf{G}$. On the other hand, the coupling strength $\varepsilon$ is not necessarily small, and the control force is purely non-invasive. However, in our case, the functions $\mathbf{f}^{(i)}$ and $\mathbf{G}^{(ij)}$ are arbitrary and $\Delta\psi^{*}_s-\Delta\psi^{*}_u$ is not necessary equal to $\pi$, but one should have small $\varepsilon$ and non-invasiveness of the control force achieved up to the order $\mathcal{O}(\varepsilon)$.

Another algorithm, which proposes similar goals as the provided unstable DFC algorithm, is called synchronization engineering and is realized experimentally in oscillatory chemical reactions~\cite{Kiss2018,Kiss2007,Rusin2009}. The synchronization engineering methodology allows one to set up the desirable coupling function, $h(\Delta \psi)$, by applying a polynomial time-delayed feedback control. Thus, in such a case, an any-phase synchronization of the two oscillators can be achieved. However, the main difference of the provided algorithm and the synchronization engineering is that here, we assume that the inter-oscillatory couplings, $\mathbf{G}^{(ij)}(\cdot,\cdot)$, naturally exist and cannot be dismissed, while in Refs.~\cite{Kiss2018,Kiss2007,Rusin2009}, it is assumed that $\mathbf{G}^{(ij)}(\cdot,\cdot)=0$ and the coupling function, $h(\Delta \psi)$, appears purely as a consequence of control signals. Therefore, we can stabilize the naturally existing unstable fixed point $\Delta\psi=\Delta\psi^{*}_u$ with the non-invasive control force, while the synchronization engineering setup requires invasive control.

\subsection{\label{subsec:two_het}Numerical demonstration of synchronization for two heterogeneous oscillators at the unstable phase difference $\Delta\psi^{*}_u$}

In this subsection, we present a numerical demonstration of the stabilization of the unstable phase difference $\Delta\psi^{*}_u$ by the non-invasive control force. We picked two different oscillators, namely FitzHugh-Nagumo (FHN) (denoted by index $i=1$) and Landau-Stuart (LS) (denoted by index $i=2$). The uncoupled FHN oscillator is described by the vector field~(\ref{fhn_f}) with the parameter $\epsilon^{(1)}\equiv \epsilon = 0.08$. It gives oscillations with the period $T^{(1)}=39.474415$. The second uncoupled oscillator is described by the vector field
\begin{equation}
\mathbf{f}^{(2)}\left(\mathbf{x}, r \right)=
\left[
\begin{array}{c}
x_{1} \left(1-x_{1}^2-x_{2}^2 \right) - \Omega^{(2)} x_{2} + r \\
x_{2} \left(1-x_{1}^2-x_{2}^2 \right) + \Omega^{(2)} x_{1}
\end{array}
\right],
\label{ls_f1}
\end{equation}
which differs from Eq.~(\ref{ls_f}) by a place where the input variable $r$ is attached. In order to satisfy weak coupling requirement $|T^{(1)}-T^{(2)}|\sim \varepsilon$, we choose the natural frequency of the LS oscillator $\Omega^{(2)}=2\pi/T^{(1)}-10^{-5} \approx 0.159161$.

\begin{figure}[h!]
	\centering\includegraphics[width=0.95\columnwidth]{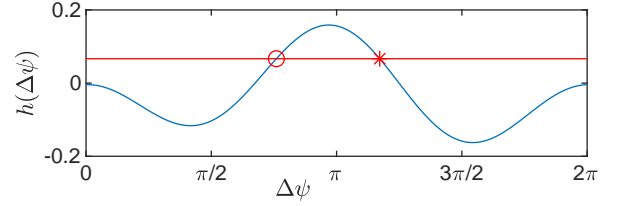}
	\caption{\label{fig:h_func} Numerically calculated function $h(\Delta \psi)$ that defines the dynamics of the phase difference $\Delta\psi$ according to Eq.~(\ref{eq:del_phi}). A horizontal straight line represents the value of the right hand side of Eq.~(\ref{eq:dpsi_su}). An asterisk and an open circle represent the stable $\Delta\psi_s^*$ and the unstable $\Delta\psi_u^*$ fixed points, respectively. As one can see, $\Delta\psi_s^*-\Delta\psi_u^* \neq \pi$ is non-standard due to heterogeneity of the oscillators and the complex inter-oscillatory couplings $\mathbf{G}^{(ij)}$.}
\end{figure}

Two coupled oscillators controlled by the unstable DFC are described by the following equations (for the index $i=1$, the index $j=2$, and vice versa):
\begin{subequations}
\label{eq:two_osc_udfcf}
\begin{align}
\dot{\mathbf{x}}^{(i)} &= \mathbf{f}^{(i)} \left(\mathbf{x}^{(i)}, r^{(i)} \right)+\varepsilon \mathbf{G}^{(ij)} \left( \mathbf{x}^{(j)},\mathbf{x}^{(i)} \right) , \label{eq:two_osc_udfcf_1} \\
\dot{w}^{(i)} &= \left( \lambda_{\mathrm{l}}^{(i)}+\lambda_{\mathrm{n}}^{(i)} s^{(i)}(t) \right) w^{(i)} \nonumber \\
&+ \kappa^{(i)}  \tilde{K}_{21}^{(i)} \left[ s^{(i)}(t-\tau(t))-s^{(i)}(t) \right] , \label{eq:two_osc_udfcf_2} \\
s^{(i)}(t) &= g^{(i)} \left(\mathbf{x}^{(i)}(t) \right), \label{eq:two_osc_udfcf_3} \\
r^{(i)}(t)&= w^{(i)}(t) .\label{eq:two_osc_udfcf_4}
\end{align}
\end{subequations}
Note that here, the time delay $\tau(t)$ is a dynamical variable that slowly depends on time as it will be described below. We set the following forms of the inter-oscillatory couplings:
\begin{equation} 
\mathbf{G}^{(12)}(\mathbf{y},\mathbf{x})= 
\left[
\begin{array}{c}
(y_{2}^2-y_1) - (x_2^2-x_1) \\
0 
\end{array}
\right]
\label{fhn_G1}
\end{equation}
and
\begin{equation} 
\mathbf{G}^{(21)}(\mathbf{x},\mathbf{y})= \frac{1}{5}
\left[
\begin{array}{c}
x_1 - y_1 \\
0 
\end{array}
\right]
.\label{ls_G1}
\end{equation}
The coupling constant $\varepsilon=1.5\times 10^{-4}$ is high enough such that without control ($r^{(1)}=r^{(2)}=0$), the coupled oscillators~(\ref{eq:two_osc_udfcf}) are in the frequency-locking regime. It can be seen from the function $h(\Delta \psi)$ defined by Eq.~(\ref{eq:h_def}) and plotted in Fig.~\ref{fig:h_func}.

The single-output functions are set to
\begin{equation} 
s^{(1)}=g^{(1)}(\mathbf{x})= x_{1}^2+x_2
\label{fhn_s1}
\end{equation}
and
\begin{equation} 
s^{(2)}=g^{(2)}(\mathbf{x})= x_{1}.
\label{ls_s1}
\end{equation}
The parameters of the DFC controller are set to the following values: $\left[\lambda_{\mathrm{l}}^{(1)},\lambda_{\mathrm{l}}^{(2)}\right]=[0.595, 0.01]$, $\left[\lambda_{\mathrm{n}}^{(1)},\lambda_{\mathrm{n}}^{(2)}\right]=[-0.25, -0.2]$, $\tilde{K}_{21}^{(1)}=\tilde{K}_{21}^{(2)}=-1$, and $\left[\kappa^{(1)},\kappa^{(2)}\right]=[4.6\times 10^{-4},0.88 \times 10^{-3}]$ such that both $\alpha^{(1)}=\alpha^{(2)}=-1$.

\begin{figure}[h!]
	\centering\includegraphics[width=0.95\columnwidth]{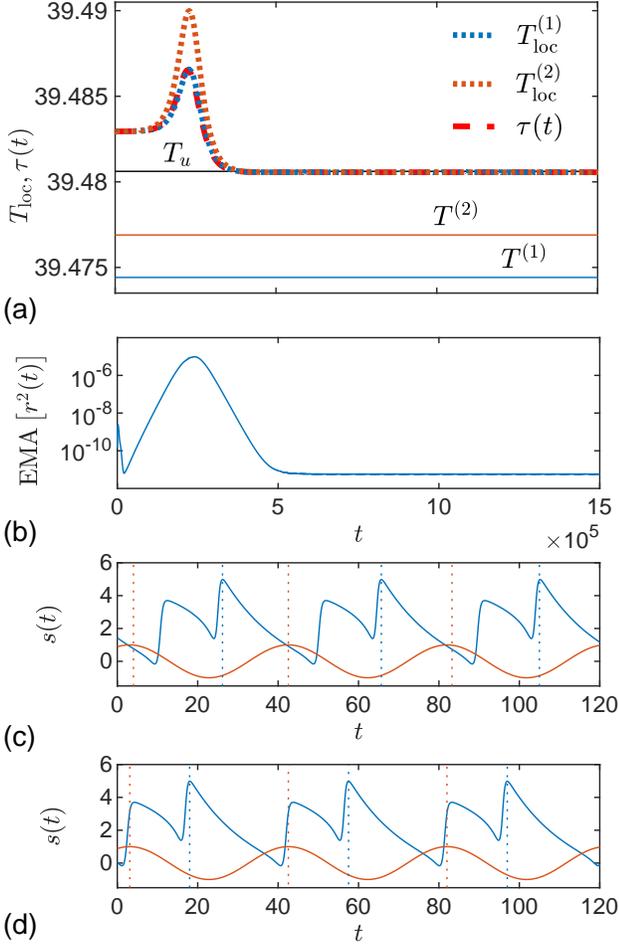}
	\caption{\label{fig:grad} The numerical simulation of the application of the unstable DFC with the adaptive  time delay to stabilize the unstable phase difference for two heterogeneous oscillators. Panel (a): the evolution of the time delay together with the evolution of ``local'' periods $T_{\mathrm{loc}}^{(i)}$ is depicted. Note that evolution of $T_{\mathrm{loc}}^{(1)}$ almost coincides with $\tau(t)$. Three horizontal lines show the values of the natural periods $T^{(i)}$ and the unstable synchronization period $T_u$. The starting point of $\tau(0)$ is equal to the stable synchronization period $T_s$, which can be obtained from Eq.~(\ref{eq:t_u}) by substituting $\Delta\psi_s^*$ to the right hand side. Panel (b): the evolution of the exponential moving average of the power of control forces, where $r^2=\left(r^{(1)}\right)^2+\left(r^{(2)}\right)^2$ and $\mathrm{EMA}\left[ \cdot\right]$ indicates the averaging procedure with an averaging window defined by $\nu$. Panel (c) and (d): the output signal from the FHN (blue color) and the LS (orange color) oscillators. Panel (c) shows a control-free regime, while panel (d) corresponds to the system under control after a transient period when $\tau$ becomes equal to $T_u$. The vertical dashed lines help to see the different phase locking values: in both panels, the orange line near $t=40$ shows time moment when the LS oscillator's output achieves maximum, while the blue lines after $t=60$ in (c) and before $t=60$ in (d) represent time moment when the FHN output is maximal.}
\end{figure}

The control scheme~(\ref{eq:two_osc_udfcf}) becomes non-invasive when the time delay is equal to $T_u$. However, $T_u$ is assumed to be unknown; therefore, we run the adaptive DFC algorithm~\cite{Pyragas2011,novi2020} to set the time delay $\tau(t)$ at the value $T_u$ by minimizing the square of the difference $W^{(1)}(t)=\left[s^{(1)}(t-\tau(t))-s^{(1)}(t)\right]^2$. Note that the adaptive algorithm in Ref.~\cite{Pyragas2011} is derived inaccurately. In particular an equation for a gradient descent method has missed factor $\alpha$. For the cases where $\alpha$ is positive, the inaccuracy can be insignificant. However, for our case of the unstable DFC scheme, $\alpha$ is always negative; thus, the algorithm provided in Ref.~\cite{Pyragas2011} gives gradient ascendance and fail to minimize the control force. Therefore, we refer to Ref.~\cite{novi2020} where corrected derivation of the adaptive DFC scheme is provided. The equations for the slowly changing time delay $\tau(t)$ read
\begin{subequations}
\label{eq:adapt}
\begin{align}
\dot{\tau} &= -\beta G , \label{eq:adapt_1} \\
\dot{G} &= 2\gamma \alpha \left[ s^{(1)}(t)-s^{(1)}\left(t-\tau(t)\right) \right] \left[ s^{(1)}(t)-u \right] -\nu G, \label{eq:adapt_2} \\
\dot{u} &= \gamma \left[ s^{(1)}(t)-u \right]. \label{eq:adapt_3}
\end{align}
\end{subequations}
Here, we have two additional dynamical variables: $G$ approximates the exponential moving average (EMA) of the derivative $\partial W^{(1)}/\partial\tau$ and $u$ is a variable for the high-pass filter. The adaptive scheme has the following parameters: $\beta$ is a speed of the gradient descent, $\nu$ determines an integration window, and $\gamma$ is a parameter for the high-pass filter. The parameters work on different time scales; thus, by selecting the values, one should keep in mind inequalities  $\gamma^{-1} < T^{(1)} < \nu^{-1} < \beta^{-1}$. Note that the proposed control scheme does not provide the optimal values for the control parameters. In the practical implementation the exact values for the control parameters, such as $\lambda_{\mathrm{l}}$, $\lambda_{\mathrm{n}}$, $\gamma$, $\nu$, $\beta$, and $\tau(0)$, should be adjusted by a trial and error method. In particular, we set the following values: $\gamma=100/T^{(1)}$, $\nu=\left( 50 T^{(1)} \right)^{-1}$, and $\beta=\nu \times 10^{-4}$.  The results from numerical simulations of Eqs.~(\ref{eq:two_osc_udfcf}) and (\ref{eq:adapt}) are depicted in Fig.~\ref{fig:grad}. As one can see in Fig.~\ref{fig:grad} (a), the adaptive time delay $\tau(t)$ successfully settled to the synchronization period $T_u$. In Fig.~\ref{fig:grad} (c) and (d), the outputs of the two oscillators are depicted for the control-free and controlled system, respectively. Figure~\ref{fig:grad}(d) shows results after a transient period when $\tau(t)$ becomes stationary. As one can see, both oscillators are frequency locked, but in (c) and (d), they have different values of $\Delta\psi$, as predicted by Fig.~\ref{fig:h_func}.

Note that strictly speaking, Eq.~(\ref{eq:adapt}) does not minimize the power of the control force $\left( r^{(1)} \right)^2+\left( r^{(2)}\right)^2$; instead, it minimizes $W^{(1)}(t)$. However, as one can see from Appendix~\ref{appsec:non_inv}, the minimization of $W^{(1)}(t)$ is equivalent to the minimization of the power of the control force.

\subsection{\label{subsec:two_ms}Numerical demonstration of the stabilization of the unstable phase difference $\Delta\psi^{*}_u$ for the master-slave coupling scheme}

In Sec.~\ref{subsec:two_het}, we analyzed the mutual coupling of the two oscillators. Here, we will investigate a master-slave configuration. In this case, the first oscillator (slave) is unidirectionally coupled to the second oscillator (master) such that only the master can affect the slave. In fact, the master oscillator is not necessarily possessing a limit cycle solution and can be any system that just injects the periodic signal to the slave oscillator.

As an illustrative example, let us imagine that there is some primitive life form that has two phases over one day: an active phase and a sleep phase. Let us say that the period of its internal biological clock is not exactly equal to 24 hours, but due to the effect of the sunlight, the internal clock synchronizes with the earth's 24-h cycle. Moreover, the active phase occurs during the day-time, while the sleep phase comes at night. Our goal would be to use the controller applied to the primitive life form such that the active phase would be at nighttime while the sleep phase comes at the day-time. Additionally, the controller should be non-invasive -- once the internal clock synchronizes with the sunlight, the control force should vanish. The possibility to do it non-invasively comes from the fact that the coupling function $h(\cdot)$ has two fixed points with alternating stability.

\begin{figure}[h!]
	\centering\includegraphics[width=0.95\columnwidth]{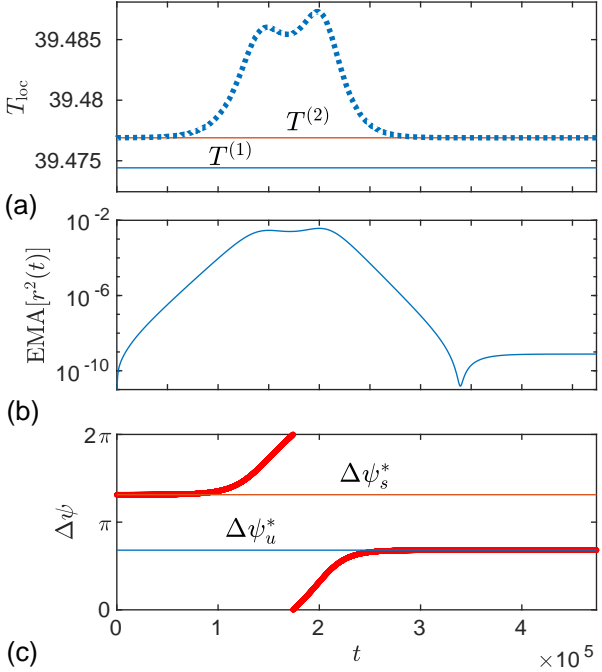}
	\caption{\label{fig:fhn_ms} The numerical demonstration of the unstable DFC usage to stabilize the unstable phase difference for the master-slave coupling configuration. Panel (a): the evolution of the ``local'' periods $T_{\mathrm{loc}}^{(1)}$ for the slave oscillator is depicted. Here, horizontal lines show the values of the natural periods $T^{(i)}$. Panel (b): the evolution of the exponential moving average of the power of the control force  applied to the slave oscillator. Here, $r \equiv r^{(1)}$, and $\mathrm{EMA}[\cdot]$ indicates an averaging procedure with the averaging window  $\nu$ defined in Sec.~\ref{subsec:two_het}. Panel (c): the evolution of the phase difference obtained as a time difference of neighboring maximums of master and slave oscillators renormalized to $2\pi$.}
\end{figure}

The derivation performed in Sec.~\ref{sec:non-inv} can be simply rewritten for the case of the master-slave scheme. By substituting $\mathbf{G}^{(21)}\left( \mathbf{x}^{(1)},\mathbf{x}^{(2)} \right)=0$, we get $H^{(21)}\left(\psi^{(1)}-\psi^{(2)}\right)=0$, and according to Eq.~(\ref{eq:h_def}), $h(\Delta\psi)=-H^{(12)}(\Delta\psi)$. The stable synchronization period, $T_s$, together with the unstable synchronization period, $T_u$, coincides with the natural period of the master oscillator $T^{(2)}$. For such a situation, the control force is applied only to the slave oscillator. We perform numerical simulations similar to that presented in Sec.~\ref{subsec:two_het} with the same values of the parameters (see Fig.~\ref{fig:fhn_ms}); yet, the LS oscillator plays the role of the master and the FHN oscillator is the slave. At the initial time moment, the slave oscillator is synchronized with the master; therefore, we assume that $T^{(2)}$ is a known parameter, and we can set the time delay $\tau=T^{(2)}$ without the need for the gradient descent procedure~(\ref{eq:adapt}). The initial phase difference $\Delta\psi=\Delta\psi_s^*$ after a transient period swaps to $\Delta\psi_u^*$ (see Fig.~\ref{fig:fhn_ms}(c)) as it was predicted by the theory. 

\section{\label{sec:netw} Disruption of synchronization in the oscillator network by applying control to a single unit}

\begin{figure}%[h!]
	\centering\includegraphics[width=0.45\columnwidth]{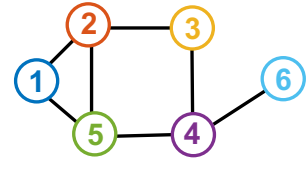}
	\caption{\label{fig:topo} Topology of the oscillator network. Different colors of the units are used to distinguish between different oscillators in Fig.~\ref{fig:netw}.}
\end{figure}

In this section, we show the effect of coupling sign inversion for the network of Landau-Stuart oscillators. Our aim is to disrupt the frequency-locking regime by applying the control only to one oscillator in the network. For this purpose we take the network with topology shown in Fig.~\ref{fig:topo}, where dynamics of each unit is described by the following equations:
\begin{equation}
\mathbf{f}^{(i)}\left(\mathbf{x}, r^{(i)} \right)=
\left[
\begin{array}{c}
x_{1} \left(1-x_{1}^2-x_{2}^2 \right) - \Omega^{(i)} x_{2} + r^{(i)} \\
x_{2} \left(1-x_{1}^2-x_{2}^2 \right) + \Omega^{(i)} x_{1}
\end{array}
\right].
\label{ls_ntw}
\end{equation}
We assume that oscillators interact through the first variables; therefore, the coupling function is
\begin{equation} 
\mathbf{G}(\mathbf{y},\mathbf{x})= 
\left[
\begin{array}{c}
2(y_{1} -x_{1}) \\
0 
\end{array}
\right]
\label{ls_G2}
\end{equation}
and the adjacency matrix elements $a^{(ij)}=1$ ($a^{(ij)}=0$) for connected (disconnected) units. As an output signal we take the first dynamical variable,
\begin{equation} 
s(t)=g(\mathbf{x}(t))= x_{1}(t). 
\label{LS_ntw}
\end{equation}

The natural frequencies of the oscillators  $\Omega^{(i)} = 2\pi/T^{(i)}$ are estimated from natural periods $T^{(i)}=2\pi+10^{-2}\times\delta T_i$ where $\delta \mathbf{T}=\left[-1.2,\, 0.4,\, 0.1,\, -0.6,\, 0.3,\, 0.8 \right]^T$. The coupling  strength is set in such a way that for the control-free system, the network is in a frequency-locking regime: $\varepsilon = 4 \times 10^{-3}$.

To gain understanding how the control affects phase dynamics of this model, one can write equations for the phase
\begin{equation}
\dot{\psi}^{(i)} = \Omega^{(i)}+\varepsilon_{\mathrm{eff}}^{(i)} \sum_{\substack{j=1 \\ j\neq i}}^{N} a^{(ij)} \sin\left( \psi^{(j)}-\psi^{(i)} \right).
\label{eq:phase_LS}
\end{equation}
As it was explained in Sec.~\ref{subs:phase_osc_contr}, the particular coupling strength can be changed by applying the control to the particular oscillator.  We choose to affect the fifth oscillator in the network since it has a large degree of connectivity. By numerically investigating Eqs.~\eqref{eq:phase_LS} with $\varepsilon_{\mathrm{eff}}^{(i)}=\varepsilon$ for $i \neq 5$ and by setting various $\varepsilon_{\mathrm{eff}}^{(5)}$ values, one can make sure that for the positive coupling values, the system remains in a synchronized regime. Even at $\varepsilon_{\mathrm{eff}}^{(5)}=0$, the network remains synchronized with the synchronization period being equal to $T^{(5)}$. The synchronization is disrupted only for negative values of $\varepsilon_{\mathrm{eff}}^{(5)}$.

\begin{figure}%[h!]
\centering\includegraphics[width=0.99\columnwidth]{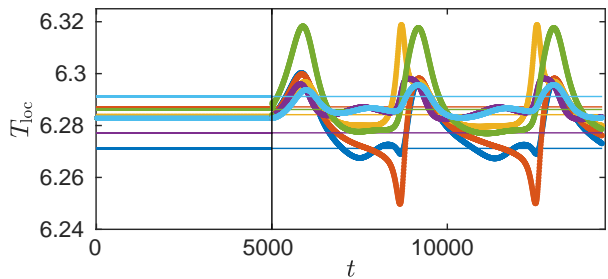}
\caption{\label{fig:netw} The dynamics of ``local'' periods of the Landau-Stuart oscillators network. Horizontal lines represent natural periods. For $t<5000$ the network is a control-free possessing frequency-locking regime, where all ``local'' periods coincide and form a horizontal line. After $t=5000$, the control is applied to the fifth oscillator. As a consequence, the ``local'' periods start to drift, and the frequency locking no longer exists. The color of symbols corresponds to the color of the oscillator in Fig.~\ref{fig:topo}.}
\end{figure}

In Fig.~\ref{fig:netw}, we demonstrate the disruption of the frequency-locking regime by applying the controller to the fifth oscillator of the Landau-Stuart oscillators network. The network is in a control-free regime until the time $t=5000$,  and after that, the controller is applied to the fifth oscillator with the following parameters: $K_{21} = -1$, $\kappa= 0.42$, $\lambda_l = 0.05$, $\lambda_n= -0.5$ and $\tau=T^{(5)}$. In such a case, $\varepsilon_{\mathrm{eff}}^{(5)}=-0.5\varepsilon$. The results confirm that the control-free system is in the frequency-locking state, while under control, the frequency locking is disrupted.

\section{\label{sec:conc}Conclusions and discussions}

In this paper, we present the unstable delayed feedback control algorithm to effectively change the sign of the coupling constant for the weakly coupled limit cycle oscillators. Since the algorithm is based on the feedback action, it works for a general class of the limit cycle oscillators possessing a single input and a single output. The controller, summarized in Eqs.~(\ref{eq:one_osc_udfcf}), contains two main parts: the unstable degree of freedom which destabilizes the oscillator, and the delayed feedback force, which returns the stability to the oscillator. The consequence of such manipulations is that the phase of the limit cycle ``feels'' the coupling constant inverted with respect to the natural one.

Numerical calculations performed in Sec.~\ref{sec:two_osc_in_anti} for the FitzHugh-Nagumo and Landau-Stuart oscillators demonstrate a successful application of the controller to switch between the in-phase and anti-phase synchronizations. Additionally, in Sec.~\ref{sec:non-inv}, we demonstrate the non-invasive nature of the controller. Two coupled oscillators or one oscillator in the presence of periodic force has two solutions for the phase difference: the stable and unstable solutions. We show that without prior knowledge of the unstable solution, the controller supplemented by the gradient descent algorithm~\cite{novi2020} is able to stabilize the unstable phase difference with vanishing control force.

The proposed algorithm is versatile for an experimental realization since an application of the algorithm does not require a priori knowledge of the structure of the oscillator and the law of mutual coupling. The possible experimental implementation of the algorithm can be realized similarly to synchronization engineering experiments presented in Refs.~\cite{Kiss2018,Kiss2007,Rusin2009}, where a collective behavior of electrochemical oscillators was controlled. The feedback signal can be constructed by electrical circuits~\cite{tamasevicius08,tam07} or by a real-time data acquisition computer.

As it is shown in Refs.~\cite{goldsztein2020antiphase} and~\cite{Pantaleone2002}, the metronomes placed on a movable platform usually synchronize in-phase, while two pendulum clocks hanging on the same beam tend to synchronize anti-phase. The provided algorithm can potentially be used to switch between both regimes. Another potential application of the algorithm can be a disruption of the synchronization in the networks. As shown in Sec.~\ref{sec:netw}, the general synchronous state is corrupted by applying the controller selectively to only one unit with a large degree of connectivity.

\appendix

\section{\label{appsec:fm_and_prc}Floquet multipliers and a phase response curve of oscillators supplemented by an unstable degree of freedom}

The $n$-dimensional oscillator supplemented by the unstable degree of freedom reads
\begin{subequations}
\label{eqapp:one_osc_udfc}
\begin{align}
\dot{\mathbf{x}} &= \mathbf{f} \left(\mathbf{x}, w \right), \label{eqapp:one_osc_udfc_1} \\
\dot{w} &= \left( \lambda_{\mathrm{l}}+\lambda_{\mathrm{n}} s(t) \right) w, \label{eqapp:one_osc_udfc_2} \\
s(t) &= g \left(\mathbf{x}(t) \right). \label{eqapp:one_osc_udfc_3}
\end{align}
\end{subequations}
By introducing notations for $n\times n$ Jacobian matrix
\begin{equation}
\mathbf{A}(t)=D_1 \mathbf{f}\left( \bm{\xi}(t) ,0 \right),
\label{eqapp:jac}
\end{equation}
and $n\times 1$ vector of the derivatives with respect to the input signal
\begin{equation}
\mathbf{p}(t)=D_2 \mathbf{f}\left( \bm{\xi}(t) ,0 \right)
\label{eqapp:p}
\end{equation}
one can write the evolution of a small perturbation near the limit cycle $(\bm{\xi}^T(t),0)$ as
\begin{equation}
\left(
\begin{array}{c}
\delta \dot{\mathbf{x}} \\
\delta \dot{w}
\end{array} \right) = \left(
\begin{array}{rl}
\mathbf{A}(t) & \mathbf{p}(t) \\
\mathbf{0}^T_n &  \lambda_{\mathrm{l}}+\lambda_{\mathrm{n}} g \left(\bm{\xi}(t) \right) \\ 
\end{array}
\right)
\left(
\begin{array}{c}
\delta \mathbf{x} \\
\delta w
\end{array} \right);
\label{eqapp:pert}
\end{equation}
here, $\mathbf{0}_n$ is an $n$-dimensional column-vector with all entries equal to zero. In order to obtain a monodromy matrix, one should solve the system~(\ref{eqapp:pert}) with $(n+1)$ different initial conditions, taken from columns of an identity matrix. Thus, for the first $n$ initial conditions, the solution for $\delta w$ is $\delta w(T)=0$. For the last solution, where $\delta w(0)=1$, one can obtain that
\begin{equation}
\delta w(T)=\exp\left[ \lambda_{\mathrm{l}}T+\lambda_{\mathrm{n}} \int_0^T g \left(\bm{\xi}(t) \right) \mathrm{d}t \right].
\label{eqapp:dw}
\end{equation}
Let us say that the control-free oscillator has $n$-dimensional monodromy matrix $\bar{\bm{\Phi}}$, which has the Floquet multipliers $(1,\mu_2,\ldots,\mu_n)$. Then, the monodromy matrix of the system~(\ref{eqapp:one_osc_udfc}) reads
\begin{equation}
\bm{\Phi}=\left(
\begin{array}{rl}
\bar{\bm{\Phi}} & \mathbf{a} \\
\mathbf{0}^T_n & \exp\left[ \lambda_{\mathrm{l}}T+\lambda_{\mathrm{n}} \int_0^T g \left(\bm{\xi}(t) \right) \mathrm{d}t \right]
\end{array}
\right);
\label{eqapp:monod}
\end{equation}
here, $\mathbf{a}$ is an $n$-dimensional column-vector containing in this context unimportant values. From the last equation, one can see that the system~(\ref{eqapp:one_osc_udfc}) possesses the same Floquet multipliers $(1,\mu_2,\ldots,\mu_n)$ as  $\bar{\bm{\Phi}}$ and one additional  $\mu_{n+1}$, defined by the right hand side of~(\ref{eqapp:dw}).

The phase response curve of the system~(\ref{eqapp:one_osc_udfc}) is a left Floquet mode corresponding to trivial Floquet multiplier $\mu_1=1$ and satisfying following equations:
\begin{equation}
\dot{\mathbf{v}}^T_1 = -\mathbf{v}^T_1 \left(
\begin{array}{rl}
\mathbf{A}(t) & \mathbf{p}(t) \\
\mathbf{0}^T_n &  \lambda_{\mathrm{l}}+\lambda_{\mathrm{n}} g \left(\bm{\xi}(t) \right) \\ 
\end{array}
\right)
.
\label{eqapp:prc}
\end{equation}
Here, the index $1$ denotes that we are looking for the first left Floquet mode corresponding to the first Floquet multiplier. The periodic solution $\mathbf{v}_1(t+T)=\mathbf{v}_1(t)$ is normalized to the first right Floquet mode,
\begin{equation}
\left[\mathbf{v}_{1,1:n}(0)\right]^T \cdot \dot{\bm{\xi}}(0)+v_{1,n+1}(0) \cdot 0=1,
\label{eqapp:prc_norm}
\end{equation}
where the additional indexes $1:n$ are used to denote a vector constructed from the vector $\mathbf{v}_1$ by taking the elements from first to $n$th. From~(\ref{eqapp:prc}) and (\ref{eqapp:prc_norm}), one can see that the first $n$ elements $\mathbf{v}_{1,1:n}(t)$ of the phase response curve of the system~(\ref{eqapp:one_osc_udfc}) coincide with the phase response curve $\bar{\mathbf{v}}_1(t)$ of the oscillator without an unstable degree of freedom, while the last component satisfies a linear non-homogeneous differential equation, 
\begin{equation}
\dot{v}_{1,n+1}=-\left[\lambda_{\mathrm{l}}+\lambda_{\mathrm{n}} g \left(\bm{\xi}(t) \right) \right] v_{1,n+1}+P(t),
\label{eqapp:prc_n_p1}
\end{equation}
with periodic function
\begin{equation}
\label{eqapp:P}
\begin{aligned}
P(t) &= \bar{\mathbf{v}}^T_1(t) \cdot \mathbf{p}(t) \\
&= \mathbf{v}^T_{1,1:n}(t) \cdot \mathbf{p}(t).
\end{aligned}
\end{equation}
The periodic solution to Eq.~(\ref{eqapp:prc_n_p1}) can be written analytically
\begin{equation}
\label{eqapp:prc_n_p1_sol}
\begin{aligned}
&v_{1,n+1}(t) =\exp\left( -\lambda_{\mathrm{l}} t- \lambda_{\mathrm{n}} \int_0^t g \left(\bm{\xi}(t^{\prime}) \right) \mathrm{d}t^{\prime} \right) \\
& \times \left[ \frac{\int_0^T P(t^{\prime}) \exp\left( \lambda_{\mathrm{l}} t^{\prime}+ \lambda_{\mathrm{n}} \int_0^{t^{\prime}} g \left(\bm{\xi}(t^{\prime\prime}) \right) \mathrm{d}t^{\prime\prime} \right) \mathrm{d}t^{\prime}}{1-\exp\left[ \lambda_{\mathrm{l}}T+\lambda_{\mathrm{n}} \int_0^T g \left(\bm{\xi}(t^{\prime}) \right) \mathrm{d}t^{\prime} \right]} \right. \\
& \left. - \int_0^t P(t^{\prime}) \exp\left( \lambda_{\mathrm{l}} t^{\prime}+ \lambda_{\mathrm{n}} \int_0^{t^{\prime}} g \left(\bm{\xi}(t^{\prime\prime}) \right) \mathrm{d}t^{\prime\prime} \right) \mathrm{d}t^{\prime} \right] .
\end{aligned}
\end{equation}

\section{\label{appsec:np1_lfm}On the $(n+1)$-th left Floquet mode}

The limit cycle oscillator supplemented by the unstable degree of freedom and governed by Eqs.~(\ref{eqapp:one_osc_udfc}) has $n+1$ left and right Floquet modes. The equation for the $i$th right Floquet mode reads
\begin{equation}
\dot{\mathbf{u}}_{i}+\Lambda_{i} \mathbf{u}_{i} = \left(
\begin{array}{rl}
\mathbf{A}(t) & \mathbf{p}(t) \\
\mathbf{0}^T_n &  \lambda_{\mathrm{l}}+\lambda_{\mathrm{n}} g \left(\bm{\xi}(t) \right) \\ 
\end{array}
\right) \mathbf{u}_{i} .
\label{eqapp:rfm}
\end{equation}
From Eq.~(\ref{eqapp:rfm}), one can see that the last component of the vector $\mathbf{u}_{i}$ is decoupled from the rest of the components
\begin{equation}
\dot{u}_{i,n+1} = \left[ \lambda_{\mathrm{l}}+\lambda_{\mathrm{n}} g \left(\bm{\xi}(t) \right)-\Lambda_{i} \right] u_{i,n+1}.
\label{eqapp:rfm1}
\end{equation}
The function $u_{i,n+1}(t)$ is $T$-periodic; thus,
\begin{equation}
u_{i,n+1}(0) = u_{i,n+1}(0) \mathrm{e}^{ \lambda_{\mathrm{l}}T+\lambda_{\mathrm{n}} \int\limits_0^T g \left(\bm{\xi}(t) \right) \mathrm{d}t-\Lambda_{i} T }.
\label{eqapp:rfm2}
\end{equation}
From the last equation one can see that for $i=1,2,\ldots,n$
\begin{equation}
u_{i,n+1}(t)=0,
\label{eqapp:rfm4}
\end{equation}
while for $i=n+1$, we have
\begin{equation}
\label{eqapp:rfm3}
\begin{aligned}
& u_{n+1,n+1}(t) = \\
& u_{n+1,n+1}(0) \mathrm{e}^{ \lambda_{\mathrm{l}}t+\lambda_{\mathrm{n}} \int\limits_0^t g \left(\bm{\xi}(t^{\prime}) \right) \mathrm{d}t^{\prime}-\Lambda_{n+1} t}.
\end{aligned}
\end{equation}
The left Floquet mode $\mathbf{v}_{n+1}$ is orthogonal to right Floquet modes $\mathbf{u}_{i}$ (where $i=1,\ldots,n$); therefore, Eq.~(\ref{eqapp:rfm4}) gives the following orthogonality relations:
\begin{equation}
\mathbf{v}^T_{n+1,1:n}(t) \cdot \mathbf{u}_{i,1:n}(t)=0.
\label{eqapp:ort}
\end{equation}
Note that $\mathbf{u}_{i,1:n}(t)$ coincides with the right Floquet modes of the oscillator without the unstable degree of freedom; thus, $\mathbf{u}_{i,1:n}(t)$ is a set of $n$ linearly independent vectors that form a full basis and, therefore, (\ref{eqapp:ort}) implies that $\mathbf{v}_{n+1,1:n}(t)=0$. While the last component with respect to Eq.~(\ref{eqapp:rfm3}) reads (without loss of generality, a normalization constant for the left Floquet mode can be chosen arbitrary; thus, one can choose $u_{n+1,n+1}(0)=1$)
\begin{equation}
\label{eqapp:lfm}
v_{n+1,n+1}(t)=  \mathrm{e}^{\Lambda_{n+1} t- \lambda_{\mathrm{l}}t - \lambda_{\mathrm{n}} \int\limits_0^t g \left(\bm{\xi}(t^{\prime}) \right) \mathrm{d}t^{\prime}}.
\end{equation}

\section{\label{appsec:np1_rfm}On the $(n+1)$-th right Floquet mode in the limit $\Lambda_{n+1}\rightarrow 0$}

In the limit $\Lambda_{n+1}\rightarrow 0$, the last component of the ${(n+1)}$-th right Floquet mode reads (cf. Eq.~(\ref{eqapp:rfm2}) and $u_{n+1,n+1}(0)=1$)
\begin{equation}
\label{eqapp:rfm5}
u_{n+1,n+1}(t) = 
\mathrm{e}^{ \lambda_{\mathrm{l}}t+\lambda_{\mathrm{n}} \int\limits_0^t g \left(\bm{\xi}(t^{\prime}) \right) \mathrm{d}t^{\prime}},
\end{equation}
while the first $n$ components, according to Eq.~(\ref{eqapp:rfm}), satisfy
\begin{equation}
\dot{\mathbf{u}}_{n+1,1:n} = \left[ \mathbf{A}(t)-\Lambda_{n+1}\mathbf{I}_{n}\right] \mathbf{u}_{n+1,1:n} + \mathbf{p}(t) u_{n+1,n+1}(t),
\label{eqapp:rfm6}
\end{equation}
where $\mathbf{I}_{n}$ is the $n$-dimensional identity matrix. Note that here, we cannot neglect small $\Lambda_{n+1}$ since the Jacobian $\mathbf{A}(t)$ gives trivial FE $\Lambda_1=0$, and interference between two FEs should be treated analytically. Equation (\ref{eqapp:rfm6}) is a linear non-homogeneous equation with periodic coefficients. First, let us write a solution to the homogeneous equation
\begin{equation}
\dot{\mathbf{u}} = \left[ \mathbf{A}(t)-\Lambda_{n+1}\mathbf{I}_{n}\right] \mathbf{u}.
\label{eqapp:hom}
\end{equation}
By denoting the evolution matrix of the oscillator without the unstable degree of  freedom as $\bar{\bm{\Phi}}(t)$, the evolution matrix to Eq.~(\ref{eqapp:hom}) is $\bm{\Phi}_{u}(t)=\bar{\bm{\Phi}}(t)\exp(-\Lambda_{n+1}t)$. Therefore the general solution to~(\ref{eqapp:rfm6}) reads
\begin{equation}
\label{eqapp:nhoms}
\begin{aligned}
&  \mathbf{u}_{n+1,1:n}(t)=  \\
& \bm{\Phi}_{u}(t) \left\lbrace \mathbf{c}+ \int\limits_0^t \bm{\Phi}^{-1}_{u}(t^{\prime}) \mathbf{p}(t^{\prime}) u_{n+1,n+1}(t^{\prime}) \mathrm{d}t^{\prime} \right\rbrace,
\end{aligned}
\end{equation}
where $\mathbf{c}$ is a vector of initial conditions. In order to have periodic vector $\mathbf{u}_{n+1,1:n}(t+T)=\mathbf{u}_{n+1,1:n}(t)$, the initial condition should satisfy
\begin{equation}
\mathbf{c} = \left[\mathbf{I}_{n}- \bm{\Phi}_{u}(T)\right]^{-1}\bm{\Phi}_{u}(T)\int\limits_0^T \bm{\Phi}^{-1}_{u}(t) \mathbf{p}(t) u_{n+1,n+1}(t) \mathrm{d}t.
\label{eqapp:init}
\end{equation}
The evolution matrix $\bar{\bm{\Phi}}(t)$ in terms of right $\bar{\mathbf{u}}_i(t)$ and left $\bar{\mathbf{v}}_i(t)$ Floquet modes of the $n$-dimensional oscillator reads
\begin{equation}
\bar{\bm{\Phi}}(t)=\bar{\mathbf{U}}(t) \mathrm{diag}\left[\exp(\Lambda_i t)\right]\bar{\mathbf{V}}(0),
\label{eqapp:evm}
\end{equation}
where $\bar{\mathbf{U}}(t)$ ($\bar{\mathbf{V}}(t)$) is a matrix filled by right (left) Floquet modes written to columns (rows) and $\mathrm{diag}\left[\exp(\Lambda_i t)\right]$ is a diagonal matrix with the value $\mathrm{e}^{\Lambda_i t}$ on the {$i$th} entry. Using (\ref{eqapp:evm}), the initial conditions Eq.~(\ref{eqapp:init}) read
\begin{equation}
\label{eqapp:init1}
\begin{aligned}
\mathbf{c} =& \bar{\mathbf{U}}(0)\mathrm{diag}\left[\left(\frac{\mu_{n+1}}{\mu_i}-1 \right)^{-1} \right] \\
& \cdot \int\limits_0^T \mathrm{diag} \left[ \mathrm{e}^{(\Lambda_{n+1}-\Lambda_{i})t} \right] \bar{\mathbf{V}}(t) \mathbf{p}(t) u_{n+1,n+1}(t) \mathrm{d}t.
\end{aligned}
\end{equation}
Finally, one can take the limit $\Lambda_{n+1}\rightarrow 0$. In the leading order of $\Lambda_{n+1}$, the matrix $\mathrm{diag}\left[\left(\mu_{n+1}/\mu_i-1 \right)^{-1} \right]$ has the first component equal to $(\Lambda_{n+1} T)^{-1}$, while the rest of the components can be neglected. Then Eq.~(\ref{eqapp:init1}) reads
\begin{equation}
\label{eqapp:init2}
\mathbf{c} = \frac{\bar{\mathbf{u}}_1(0)}{\Lambda_{n+1}T} \int\limits_0^T  \bar{\mathbf{v}}^T_1(t) \mathbf{p}(t) u_{n+1,n+1}(t) \mathrm{d}t.
\end{equation}
The last equation shows that $\mathbf{c}$ goes to infinity in the limit $\Lambda_{n+1}\rightarrow 0$, while the second term in curly brackets of Eq.~(\ref{eqapp:nhoms}) remains finite and thus can be ignored. Then Eq.~(\ref{eqapp:nhoms}) simplifies to
\begin{equation}
\label{eqapp:nhoms2}
\mathbf{u}_{n+1,1:n}(t)= \frac{\bar{\mathbf{u}}_1(t)}{\Lambda_{n+1}T} \int\limits_0^T  \bar{\mathbf{v}}^T_1(t^{\prime}) \mathbf{p}(t^{\prime}) u_{n+1,n+1}(t^{\prime}) \mathrm{d}t^{\prime}.
\end{equation}
Using (\ref{eqapp:rfm5}) and recalling that $\bar{\mathbf{v}}_1(t)=\mathbf{v}_{1,1:n}(t)$, Eq.~(\ref{eqapp:nhoms2}) reads
\begin{equation}
\label{eqapp:nhoms3}
\begin{aligned}
& \mathbf{u}_{n+1,1:n}(t)=  \frac{\dot{\bm{\xi}}(t)}{\Lambda_{n+1}T} \\
& \int\limits_0^T  \mathbf{v}^T_{1,1:n}(t^{\prime}) \mathbf{p}(t^{\prime}) \exp \left[ \lambda_{\mathrm{l}}t^{\prime}+\lambda_{\mathrm{n}} \int\limits_0^{t^{\prime}} g \left(\bm{\xi}(t^{\prime\prime}) \right) \mathrm{d}t^{\prime\prime} \right] \mathrm{d}t^{\prime}.
\end{aligned}
\end{equation}

\section{\label{appsec:add_per_orbit}Analytical expression of an additional periodic orbit}

According to Sec.~\ref{sec:udfc_oscillator}, the dynamical system under the DFC described by Eq.~(\ref{eq:one_osc_udfcf}) with $\tau=T$ has the periodic solution $(\mathbf{x}^T (t),w(t))=(\bm{\xi}^T (t),0)$ for any value of $\kappa$. At point $\kappa=\kappa^{*}$, the limit cycle $(\bm{\xi}^T (t),0)$ has two degenerate FMs $\mu=1$; therefore, there should be two linearly independent real Floquet modes corresponding to the FMs. One of them is the trivial Floquet mode represented as a derivative of the limit cycle $\mathbf{u}_{1}(t)=(\dot{\bm{\xi}}^T(t),0)^T$. According to Eqs.~(\ref{eq:dp_relation}), all real Floquet modes of the DFC system can be mapped to the PFC system and vice versa; however, if we substitute $\kappa=\kappa^{*}$ and $\Lambda_D(\kappa^{*})=0$, we get $\Lambda_{P}=0$ and therefore, $\Gamma=\kappa^{*}[1-\exp(-\Lambda_P T)]=0$. At $\Gamma=0$, the PFC system has only one Floquet mode corresponding to $\mu=1$, that is, the trivial Floquet mode. The reason for such incompatibility is that for the DFC system, an algebraic multiplicity of the eigenvalue $\mu=1$ is equal to $2$, while a geometric multiplicity is equal to $1$, and we should look for the second Floquet mode as being a generalized eigen-vector of a monodromy operator rather than a regular eigen-vector of a monodromy operator. Such a generalized Floquet mode, denoted as $\mathbf{u}_{1,\mathrm{gen}}(t+T)=\mathbf{u}_{1,\mathrm{gen}}(t)$, corresponds to an additional periodic orbit that coalesces with the limit cycle $(\bm{\xi}^T (t),0)$ at $\kappa=\kappa^{*}$. The point $\kappa=\kappa^{*}$ is the transcritical bifurcation point, and the generalized Floquet mode $\mathbf{u}_{1,\mathrm{gen}}(t)$ is nothing more than
\begin{equation}
\label{eqapp:gen_fm}
\mathbf{u}_{1,\mathrm{gen}}(t)=\frac{\mathbf{u}^{(1)}_{\mathrm{P},1}(t)}{\Lambda^{\prime}_{\mathrm{P},1}(0) T},
\end{equation}
where $\mathbf{u}^{(1)}_{\mathrm{P},1}(t)$ is defined by Eq.~(\ref{eq:fm_tr}).

Let us analyze small perturbations from the limit cycle $(\bm{\xi}^T (t),0)$ along Floquet modes. Starting from an initial state $(\bm{\xi}^T (t),0)^T+ \Delta \mathbf{u}_{1}(t)$, where $\Delta$ is infinitely small quantity having a time dimension, after evolution over period $T$ (equivalently one can say that after acting with the monodromy operator), the final state will be exactly the same initial state because $\mathbf{u}_{1}(t)$ represents perturbation along the limit cycle or in other words, $\mathbf{u}_{1}(t)$ is an eigen-vector of the monodromy operator with the eigenvalue equal to $1$. Next, let us consider the initial state $(\bm{\xi}^T (t),0)^T+ \Delta \mathbf{u}_{1,\mathrm{gen}}(t)$. After acting on it by the monodromy operator (note that $\mathbf{u}_{1,\mathrm{gen}}(t)$ is a generalized eigen-vector of rank $2$), we will get $(\bm{\xi}^T (t),0)^T+ \Delta \mathbf{u}_{1,\mathrm{gen}}(t)+\Delta \mathbf{u}_{1}(t)$, and if we additionally evolve such state backward in time by the small amount $\Delta$, we will end up with the same initial state $(\bm{\xi}^T (t),0)^T+ \Delta \mathbf{u}_{1,\mathrm{gen}}(t)$. Since we performed evolution over time $T_{\mathrm{add}}=T-\Delta$ and returned back to an initial state, one can say that there is additional periodic solution
\begin{equation}
\label{eqapp:add_per}
\left(
\begin{array}{c}
\bm{\xi}_{\mathrm{add}}(t) \\
w_{\mathrm{add}}(t)  \\ 
\end{array}
\right)
=\left(
\begin{array}{c}
\bm{\xi}\left(t\frac{T}{T_{\mathrm{add}}}\right) \\
0  \\ 
\end{array}
\right)+\Delta \mathbf{u}_{1,\mathrm{gen}}\left( t\frac{T}{T_{\mathrm{add}}}\right)
\end{equation}
with the period $T_{\mathrm{add}}$. Note that the existence of an additional periodic solution~(\ref{eqapp:add_per}) agrees with Ref.~\cite{hoo12} and~\cite{Pyragas2013}. In order to check that Eq.~(\ref{eqapp:add_per}) is a solution of the system~(\ref{eq:one_osc_udfcf}), let us set $\kappa=\kappa^{*}+\Delta_{\kappa}$, where $\Delta_{\kappa}\equiv \Delta_{\kappa}(\Delta)\sim \mathcal{O}(\Delta)$ is a small deviation from the threshold value $\kappa^{*}$, then insert (\ref{eqapp:add_per}) into both sides of Eqs.~(\ref{eq:one_osc_udfcf}) and collect the terms up to the order $\mathcal{O}(\Delta)$. For the convenience, let us rewrite (\ref{eqapp:add_per}) as
\begin{equation}
\label{eqapp:add_per1}
\begin{aligned}
\left(
\begin{array}{c}
\bm{\xi}_{\mathrm{add}}(t) \\
w_{\mathrm{add}}(t)  \\ 
\end{array}
\right)
=& \left(
\begin{array}{c}
\bm{\xi}\left(t\left(1+\frac{\Delta}{T}\right)\right) \\
0  \\ 
\end{array}
\right) \\
&+\frac{\Delta}{\Lambda^{\prime}_{\mathrm{P},1}(0) T} \mathbf{u}^{(1)}_{\mathrm{P},1}\left(t\left(1+\frac{\Delta}{T}\right)\right)
\end{aligned}
\end{equation}
Now, by substituting~(\ref{eqapp:add_per1}) into the left hand side of Eqs.~(\ref{eq:one_osc_udfcf}), one gets (here $t^{\prime}=t(1+\Delta/T)$)
\begin{equation}
\label{eqapp:add_per_der}
\left(
\begin{array}{c}
\dot{\bm{\xi}}(t^{\prime}) \\
0  \\ 
\end{array}
\right) \left(1+\frac{\Delta}{T} \right)+\frac{\Delta}{\Lambda^{\prime}_{\mathrm{P},1}(0) T} \dot{\mathbf{u}}^{(1)}_{\mathrm{P},1}(t^{\prime}),
\end{equation}
while substituting~(\ref{eqapp:add_per1}) into the right hand side of Eqs.~(\ref{eq:one_osc_udfcf}), one can see that $\Delta_{\kappa}$ appears multiplied by $\Delta$; thus, it can be dismissed; therefore, one gets
\begin{equation}
\label{eqapp:add_per_rhs}
\begin{aligned}
&\left(
\begin{array}{c}
\mathbf{f}(\bm{\xi}(t^{\prime}),0) \\
0  \\ 
\end{array}
\right) \\
&+ \frac{\Delta}{\Lambda^{\prime}_{\mathrm{P},1}(0) T} \left(
  \begin{array}{rl}
  \mathbf{A}(t^{\prime}) & \mathbf{p}(t^{\prime}) \\
  \mathbf{0}^T_n & \lambda_{\mathrm{l}}+\lambda_{\mathrm{n}} g(\bm{\xi}(t^{\prime}))
  \end{array}
     \right)\mathbf{u}^{(1)}_{\mathrm{P},1}(t^{\prime}) \\
& - \kappa^{*}\Delta
\left(
  \begin{array}{rl}
  \mathbf{0}_{n\times n} & \mathbf{0}_n \\
  \tilde{K}_{21}\mathbf{q}^T(t^{\prime}) & 0
  \end{array}
     \right)
\left(
\begin{array}{c}
\dot{\bm{\xi}}(t^{\prime}) \\
0  \\ 
\end{array}
\right).
\end{aligned}
\end{equation}
From~(\ref{eq:kappa_tr1}), one can see that $\kappa^{*}=1/\left(\Lambda^{\prime}_{\mathrm{P},1}(0) T\right)$; therefore, by equating (\ref{eqapp:add_per_der}) and (\ref{eqapp:add_per_rhs}), one can prove that (\ref{eqapp:add_per}) indeed is a periodic solution with the period $T_{\mathrm{add}}$ and $\mathbf{u}^{(1)}_{\mathrm{P},1}(t)$ satisfies Eq.~(\ref{eq:fm_pfc_first}).

\section{\label{appsec:non_inv}$\mathcal{O}(\varepsilon)$ order non-invasiveness of the control force for two coupled oscillators}

First, let us consider the control-free oscillators Eqs.~(\ref{eq:two_contr_free}). The phase reduction allows us to obtain phase dynamics up to the order $\mathcal{O}(\varepsilon)$. On the other hand, in order to obtain the state of the oscillators $\mathbf{x}^{(i)}(t)$ up to the order $\mathcal{O}(\varepsilon)$, one should perform additional calculations. The phase model~(\ref{eq:del_phi}) possesses the solution $\Delta\psi^{*}_u$; therefore, the state of the oscillators can be written in the form
\begin{subequations}
\label{eqapp:st_expan}
\begin{align}
\mathbf{x}^{(1)}(t) &= \bm{\xi}^{(1)}\left(t\frac{T^{(1)}}{T_u} \right)+\varepsilon \bm{\zeta}^{(1)}(t)+\mathcal{O}\left(\varepsilon^2 \right), \label{eqapp:st_expan_1} \\
\mathbf{x}^{(2)}(t) &= \bm{\xi}^{(2)}\left(t\frac{T^{(2)}}{T_u} +\frac{\Delta\psi^{*}_u}{2\pi}T^{(2)}\right)+\varepsilon \bm{\zeta}^{(2)}(t)+\mathcal{O}\left(\varepsilon^2 \right), \label{eqapp:st_expan_2}
\end{align}
\end{subequations}
where $T_u$ is the synchronization period obtained by~(\ref{eq:t_u}), $\bm{\xi}^{(i)}\left(t+T^{(i)} \right)=\bm{\xi}^{(i)}\left(t \right)$ is a periodic solution to uncoupled oscillators ($\varepsilon=0$) that plays the role of the zeroth-order term $\mathcal{O}(\varepsilon^0)$ for the expansion (\ref{eqapp:st_expan}), and the periodic function $\varepsilon\bm{\zeta}^{(i)}\left(t+T_u \right)=\varepsilon\bm{\zeta}^{(i)}\left(t\right)$ is a first-order correction to the oscillator's state vector. Note that further expansion (say, $\mathcal{O}(\varepsilon^2)$-order terms and higher) cannot be performed without supplement of the phase model~(\ref{eq:del_phi}) by additional $\mathcal{O}(\varepsilon^j)$-order terms.

The next step is to obtain the functions $\bm{\zeta}^{(i)}(t)$. We put the expansion~(\ref{eqapp:st_expan}) into Eqs.~(\ref{eq:two_contr_free}) and collect $\mathcal{O}(\varepsilon)$-order terms on both sides of the equations (note that here, we use Eq.~(\ref{eq:t_u}) in order to approximate $T^{(1)}/T_u \approx 1+\frac{\varepsilon}{\Omega} H^{(12)}\left(\Delta\psi^{*}_u \right)$ and $T^{(2)}/T_u \approx 1+\frac{\varepsilon}{\Omega} H^{(21)}\left(-\Delta\psi^{*}_u \right)$),
\begin{subequations}
\label{eqapp:zeta}
\begin{align}
\dot{\bm{\zeta}}^{(1)}(t) &= \mathbf{A}^{(1)}\left(t\frac{T^{(1)}}{T_u} \right) \bm{\zeta}^{(1)}(t) \nonumber \\
&+ \mathbf{G}^{(12)}\left(  \bm{\xi}^{(2)}\left(t\frac{T^{(2)}}{T_u} +\frac{\Delta\psi^{*}_u}{2\pi}T^{(2)}\right),\bm{\xi}^{(1)}\left(t\frac{T^{(1)}}{T_u} \right)\right) \nonumber \\
&-\frac{H^{(12)}\left(\Delta\psi^{*}_u \right)}{\Omega} \dot{\bm{\xi}}^{(1)}\left(t\frac{T^{(1)}}{T_u} \right) , \label{eqapp:zeta_1} \\
\dot{\bm{\zeta}}^{(2)}(t) &= \mathbf{A}^{(2)}\left(t\frac{T^{(2)}}{T_u} +\frac{\Delta\psi^{*}_u}{2\pi}T^{(2)}\right) \bm{\zeta}^{(2)}(t) \nonumber \\
&+ \mathbf{G}^{(21)}\left( \bm{\xi}^{(1)}\left(t\frac{T^{(1)}}{T_u} \right) ,\bm{\xi}^{(2)}\left(t\frac{T^{(2)}}{T_u} +\frac{\Delta\psi^{*}_u}{2\pi}T^{(2)}\right)\right) \nonumber \\
&-\frac{H^{(21)}\left(-\Delta\psi^{*}_u \right)}{\Omega} \dot{\bm{\xi}}^{(2)}\left(t\frac{T^{(2)}}{T_u} +\frac{\Delta\psi^{*}_u}{2\pi}T^{(2)}\right) , \label{eqapp:zeta_2}
\end{align}
\end{subequations}
where $\mathbf{A}^{(i)}(\cdot)$ is the Jacobian of the $i$th oscillator defined by~(\ref{eq:jac}). Although Eqs.~(\ref{eqapp:zeta}) contains an undefined frequency of the ``central'' oscillator $\Omega$, in order to avoid uncertainties without loss of the accuracy, one can substitute, for example, $\Omega=\Omega^{(1)}$ to~(\ref{eqapp:zeta_1}) and $\Omega=\Omega^{(2)}$ to~(\ref{eqapp:zeta_2}). The last equations are first-order linear non-homogeneous differential equations with periodic coefficients, and its periodic solution can be obtained by using Floquet theory~\cite{Chicone2006}. However, as we will see later, we do not need to have explicit solutions; instead, it is enough to have an explicit differential equation~(\ref{eqapp:zeta}).

Now, let us consider two oscillator's system~(\ref{eq:two_contr}) under the unstable DFC described by Eqs.~(\ref{eq:one_osc_udfcf}) with $\tau^{(1)}=\tau^{(2)}=T_u$. Similar to Eq.~(\ref{eqapp:st_expan}), one should perform the expansion of the state vectors $\left( \left[\mathbf{x}^{(i)}(t) \right]^T,w^{(i)}(t) \right)^T$ up to the $\mathcal{O}(\varepsilon^2)$-order and show that the $\mathcal{O}(\varepsilon)$-order term for the additional variable $w^{(i)}(t)$ is zero. Fortunately, it is very easy to do. In fact, one can check that the form
\begin{subequations}
\label{eqapp:st_expan_add}
\begin{align}
\left(
\begin{array}{c}
\mathbf{x}^{(1)}(t) \\
w^{(1)}(t)
\end{array}
\right)
 &= 
\left(
\begin{array}{c}
 \bm{\xi}^{(1)}\left(t\frac{T^{(1)}}{T_u} \right) \\
 0
\end{array}
\right)
 +\varepsilon
 \left(
\begin{array}{c}
\bm{\zeta}^{(1)}(t) \\
0
\end{array}
\right)
 +\mathcal{O}\left(\varepsilon^2 \right), \label{eqapp:st_expan_add_1} \\
\left(
\begin{array}{c}
\mathbf{x}^{(2)}(t) \\
w^{(2)}(t)
\end{array}
\right)
 &= \left(
 \begin{array}{c}
 \bm{\xi}^{(2)}\left(t\frac{T^{(2)}}{T_u} +\frac{\Delta\psi^{*}_u}{2\pi}T^{(2)}\right) \\
 0
 \end{array}
  \right)
 +\varepsilon
 \left(
\begin{array}{c}
\bm{\zeta}^{(2)}(t) \\
0
\end{array}
\right) \nonumber \\
 &+\mathcal{O}\left(\varepsilon^2 \right), \label{eqapp:st_expan_add_2}
\end{align}
\end{subequations}
is a solution to~(\ref{eq:two_contr}) where the functions $\bm{\zeta}^{(i)}(t)$ are the same functions defined by Eqs.~(\ref{eqapp:zeta}). Thus, we conclude that the $\mathcal{O}(\varepsilon)$-order term for the variable $w^{(i)}(t)$ is zero. Since the control force $r^{(i)}(t)=w^{(i)}(t)$, we end up with $\mathcal{O}(\varepsilon)$-order non-invasiveness of the control force.

\section*{Acknowledgments}
Authors would like to thank Thomas Gajdosik, Egidijus Anisimovas and Lukas Razinkovas for comments.

\section*{Data Availability}
The data that support the findings of this study are available from the corresponding author upon reasonable request.

\bibliography{references}% Produces the bibliography via BibTeX.

%aipnum4-2.bst 2019-01-14 (MD) hand-edited version of apsrev4-1.bst
\begin{thebibliography}{37}%
\makeatletter
\providecommand \@ifxundefined [1]{%
 \@ifx{#1\undefined}
}%
\providecommand \@ifnum [1]{%
 \ifnum #1\expandafter \@firstoftwo
 \else \expandafter \@secondoftwo
 \fi
}%
\providecommand \@ifx [1]{%
 \ifx #1\expandafter \@firstoftwo
 \else \expandafter \@secondoftwo
 \fi
}%
\providecommand \natexlab [1]{#1}%
\providecommand \enquote  [1]{``#1''}%
\providecommand \bibnamefont  [1]{#1}%
\providecommand \bibfnamefont [1]{#1}%
\providecommand \citenamefont [1]{#1}%
\providecommand \href@noop [0]{\@secondoftwo}%
\providecommand \href [0]{\begingroup \@sanitize@url \@href}%
\providecommand \@href[1]{\@@startlink{#1}\@@href}%
\providecommand \@@href[1]{\endgroup#1\@@endlink}%
\providecommand \@sanitize@url [0]{\catcode `\\12\catcode `\$12\catcode
  `\&12\catcode `\#12\catcode `\^12\catcode `\_12\catcode `\%12\relax}%
\providecommand \@@startlink[1]{}%
\providecommand \@@endlink[0]{}%
\providecommand \url  [0]{\begingroup\@sanitize@url \@url }%
\providecommand \@url [1]{\endgroup\@href {#1}{\urlprefix }}%
\providecommand \urlprefix  [0]{URL }%
\providecommand \Eprint [0]{\href }%
\providecommand \doibase [0]{https://doi.org/}%
\providecommand \selectlanguage [0]{\@gobble}%
\providecommand \bibinfo  [0]{\@secondoftwo}%
\providecommand \bibfield  [0]{\@secondoftwo}%
\providecommand \translation [1]{[#1]}%
\providecommand \BibitemOpen [0]{}%
\providecommand \bibitemStop [0]{}%
\providecommand \bibitemNoStop [0]{.\EOS\space}%
\providecommand \EOS [0]{\spacefactor3000\relax}%
\providecommand \BibitemShut  [1]{\csname bibitem#1\endcsname}%
\let\auto@bib@innerbib\@empty
%</preamble>
\bibitem [{\citenamefont {Buck}(1988)}]{Buck1988}%
  \BibitemOpen
  \bibfield  {author} {\bibinfo {author} {\bibfnamefont {J.}~\bibnamefont
  {Buck}},\ }\bibfield  {title} {\enquote {\bibinfo {title} {Synchronous
  rhythmic flashing of fireflies. ii.}}\ }\href@noop {} {\bibfield  {journal}
  {\bibinfo  {journal} {Q. Rev. Biol.}\ }\textbf {\bibinfo {volume} {63}},\
  \bibinfo {pages} {265--289} (\bibinfo {year} {1988})}\BibitemShut {NoStop}%
\bibitem [{\citenamefont {Friedrich}\ and\ \citenamefont
  {J\"ulicher}(2012)}]{Friedrich12}%
  \BibitemOpen
  \bibfield  {author} {\bibinfo {author} {\bibfnamefont {B.~M.}\ \bibnamefont
  {Friedrich}}\ and\ \bibinfo {author} {\bibfnamefont {F.}~\bibnamefont
  {J\"ulicher}},\ }\bibfield  {title} {\enquote {\bibinfo {title} {Flagellar
  synchronization independent of hydrodynamic interactions},}\ }\href
  {https://doi.org/10.1103/PhysRevLett.109.138102} {\bibfield  {journal}
  {\bibinfo  {journal} {Phys. Rev. Lett.}\ }\textbf {\bibinfo {volume} {109}},\
  \bibinfo {pages} {138102} (\bibinfo {year} {2012})}\BibitemShut {NoStop}%
\bibitem [{\citenamefont {Klindt}\ \emph {et~al.}(2017)\citenamefont {Klindt},
  \citenamefont {Ruloff}, \citenamefont {Wagner},\ and\ \citenamefont
  {Friedrich}}]{Klindt2017}%
  \BibitemOpen
  \bibfield  {author} {\bibinfo {author} {\bibfnamefont {G.~S.}\ \bibnamefont
  {Klindt}}, \bibinfo {author} {\bibfnamefont {C.}~\bibnamefont {Ruloff}},
  \bibinfo {author} {\bibfnamefont {C.}~\bibnamefont {Wagner}},\ and\ \bibinfo
  {author} {\bibfnamefont {B.~M.}\ \bibnamefont {Friedrich}},\ }\bibfield
  {title} {\enquote {\bibinfo {title} {In-phase and anti-phase flagellar
  synchronization by waveform compliance and basal coupling},}\ }\href
  {http://stacks.iop.org/1367-2630/19/i=11/a=113052} {\bibfield  {journal}
  {\bibinfo  {journal} {New Journal of Physics}\ }\textbf {\bibinfo {volume}
  {19}},\ \bibinfo {pages} {113052} (\bibinfo {year} {2017})}\BibitemShut
  {NoStop}%
\bibitem [{\citenamefont {Nakao}, \citenamefont {Yanagita},\ and\ \citenamefont
  {Kawamura}(2014)}]{nakao14}%
  \BibitemOpen
  \bibfield  {author} {\bibinfo {author} {\bibfnamefont {H.}~\bibnamefont
  {Nakao}}, \bibinfo {author} {\bibfnamefont {T.}~\bibnamefont {Yanagita}},\
  and\ \bibinfo {author} {\bibfnamefont {Y.}~\bibnamefont {Kawamura}},\
  }\bibfield  {title} {\enquote {\bibinfo {title} {Phase-reduction approach to
  synchronization of spatiotemporal rhythms in reaction-diffusion systems},}\
  }\href {https://doi.org/10.1103/PhysRevX.4.021032} {\bibfield  {journal}
  {\bibinfo  {journal} {Phys. Rev. X}\ }\textbf {\bibinfo {volume} {4}},\
  \bibinfo {pages} {021032} (\bibinfo {year} {2014})}\BibitemShut {NoStop}%
\bibitem [{\citenamefont {Kiss}, \citenamefont {Zhai},\ and\ \citenamefont
  {Hudson}(2002)}]{Kiss2002}%
  \BibitemOpen
  \bibfield  {author} {\bibinfo {author} {\bibfnamefont {I.~Z.}\ \bibnamefont
  {Kiss}}, \bibinfo {author} {\bibfnamefont {Y.}~\bibnamefont {Zhai}},\ and\
  \bibinfo {author} {\bibfnamefont {J.~L.}\ \bibnamefont {Hudson}},\ }\bibfield
   {title} {\enquote {\bibinfo {title} {Emerging coherence in a population of
  chemical oscillators},}\ }\href {https://doi.org/10.1126/science.1070757}
  {\bibfield  {journal} {\bibinfo  {journal} {Science}\ }\textbf {\bibinfo
  {volume} {296}},\ \bibinfo {pages} {1676--1678} (\bibinfo {year} {2002})},\
  \Eprint
  {https://arxiv.org/abs/http://www.sciencemag.org/content/296/5573/1676.full.pdf}
  {http://www.sciencemag.org/content/296/5573/1676.full.pdf} \BibitemShut
  {NoStop}%
\bibitem [{\citenamefont {Wiesenfeld}, \citenamefont {Colet},\ and\
  \citenamefont {Strogatz}(1998)}]{PhysRevE.57.1563}%
  \BibitemOpen
  \bibfield  {author} {\bibinfo {author} {\bibfnamefont {K.}~\bibnamefont
  {Wiesenfeld}}, \bibinfo {author} {\bibfnamefont {P.}~\bibnamefont {Colet}},\
  and\ \bibinfo {author} {\bibfnamefont {S.~H.}\ \bibnamefont {Strogatz}},\
  }\bibfield  {title} {\enquote {\bibinfo {title} {Frequency locking in
  {J}osephson arrays: Connection with the kuramoto model},}\ }\href
  {https://doi.org/10.1103/PhysRevE.57.1563} {\bibfield  {journal} {\bibinfo
  {journal} {Phys. Rev. E}\ }\textbf {\bibinfo {volume} {57}},\ \bibinfo
  {pages} {1563--1569} (\bibinfo {year} {1998})}\BibitemShut {NoStop}%
\bibitem [{\citenamefont {Weiss}, \citenamefont {Kronwald},\ and\ \citenamefont
  {Marquardt}(2016)}]{Weiss2016}%
  \BibitemOpen
  \bibfield  {author} {\bibinfo {author} {\bibfnamefont {T.}~\bibnamefont
  {Weiss}}, \bibinfo {author} {\bibfnamefont {A.}~\bibnamefont {Kronwald}},\
  and\ \bibinfo {author} {\bibfnamefont {F.}~\bibnamefont {Marquardt}},\
  }\bibfield  {title} {\enquote {\bibinfo {title} {Noise-induced transitions in
  optomechanical synchronization},}\ }\href
  {http://stacks.iop.org/1367-2630/18/i=1/a=013043} {\bibfield  {journal}
  {\bibinfo  {journal} {New Journal of Physics}\ }\textbf {\bibinfo {volume}
  {18}},\ \bibinfo {pages} {013043} (\bibinfo {year} {2016})}\BibitemShut
  {NoStop}%
\bibitem [{\citenamefont {N\'eda}\ \emph
  {et~al.}(2000{\natexlab{a}})\citenamefont {N\'eda}, \citenamefont {Ravasz},
  \citenamefont {Brechet}, \citenamefont {Vicsek},\ and\ \citenamefont
  {Barab\'asi}}]{Neda2000}%
  \BibitemOpen
  \bibfield  {author} {\bibinfo {author} {\bibfnamefont {Z.}~\bibnamefont
  {N\'eda}}, \bibinfo {author} {\bibfnamefont {E.}~\bibnamefont {Ravasz}},
  \bibinfo {author} {\bibfnamefont {Y.}~\bibnamefont {Brechet}}, \bibinfo
  {author} {\bibfnamefont {T.}~\bibnamefont {Vicsek}},\ and\ \bibinfo {author}
  {\bibfnamefont {A.-L.}\ \bibnamefont {Barab\'asi}},\ }\bibfield  {title}
  {\enquote {\bibinfo {title} {The sound of many hands clapping},}\ }\href
  {http://dx.doi.org/10.1038/35002660} {\bibfield  {journal} {\bibinfo
  {journal} {Nature}\ }\textbf {\bibinfo {volume} {403}},\ \bibinfo {pages}
  {849--} (\bibinfo {year} {2000}{\natexlab{a}})}\BibitemShut {NoStop}%
\bibitem [{\citenamefont {N\'eda}\ \emph
  {et~al.}(2000{\natexlab{b}})\citenamefont {N\'eda}, \citenamefont {Ravasz},
  \citenamefont {Vicsek}, \citenamefont {Brechet},\ and\ \citenamefont
  {Barab\'asi}}]{Neda2000a}%
  \BibitemOpen
  \bibfield  {author} {\bibinfo {author} {\bibfnamefont {Z.}~\bibnamefont
  {N\'eda}}, \bibinfo {author} {\bibfnamefont {E.}~\bibnamefont {Ravasz}},
  \bibinfo {author} {\bibfnamefont {T.}~\bibnamefont {Vicsek}}, \bibinfo
  {author} {\bibfnamefont {Y.}~\bibnamefont {Brechet}},\ and\ \bibinfo {author}
  {\bibfnamefont {A.~L.}\ \bibnamefont {Barab\'asi}},\ }\bibfield  {title}
  {\enquote {\bibinfo {title} {Physics of the rhythmic applause},}\ }\href
  {https://doi.org/10.1103/PhysRevE.61.6987} {\bibfield  {journal} {\bibinfo
  {journal} {Phys. Rev. E}\ }\textbf {\bibinfo {volume} {61}},\ \bibinfo
  {pages} {6987--6992} (\bibinfo {year} {2000}{\natexlab{b}})}\BibitemShut
  {NoStop}%
\bibitem [{\citenamefont {Motter}\ \emph {et~al.}(2013)\citenamefont {Motter},
  \citenamefont {Myers}, \citenamefont {Anghel},\ and\ \citenamefont
  {Nishikawa}}]{Motter2013}%
  \BibitemOpen
  \bibfield  {author} {\bibinfo {author} {\bibfnamefont {A.~E.}\ \bibnamefont
  {Motter}}, \bibinfo {author} {\bibfnamefont {S.~A.}\ \bibnamefont {Myers}},
  \bibinfo {author} {\bibfnamefont {M.}~\bibnamefont {Anghel}},\ and\ \bibinfo
  {author} {\bibfnamefont {T.}~\bibnamefont {Nishikawa}},\ }\bibfield  {title}
  {\enquote {\bibinfo {title} {Spontaneous synchrony in power-grid networks},}\
  }\href {http://dx.doi.org/10.1038/nphys2535} {\bibfield  {journal} {\bibinfo
  {journal} {Nat Phys}\ }\textbf {\bibinfo {volume} {9}},\ \bibinfo {pages}
  {191--197} (\bibinfo {year} {2013})}\BibitemShut {NoStop}%
\bibitem [{\citenamefont {D\"{o}rfler}, \citenamefont {Chertkov},\ and\
  \citenamefont {Bullo}(2013)}]{Dorfler2013}%
  \BibitemOpen
  \bibfield  {author} {\bibinfo {author} {\bibfnamefont {F.}~\bibnamefont
  {D\"{o}rfler}}, \bibinfo {author} {\bibfnamefont {M.}~\bibnamefont
  {Chertkov}},\ and\ \bibinfo {author} {\bibfnamefont {F.}~\bibnamefont
  {Bullo}},\ }\bibfield  {title} {\enquote {\bibinfo {title} {Synchronization
  in complex oscillator networks and smart grids},}\ }\href
  {https://doi.org/10.1073/pnas.1212134110} {\bibfield  {journal} {\bibinfo
  {journal} {Proceedings of the National Academy of Sciences}\ }\textbf
  {\bibinfo {volume} {110}},\ \bibinfo {pages} {2005--2010} (\bibinfo {year}
  {2013})},\ \Eprint
  {https://arxiv.org/abs/http://www.pnas.org/content/110/6/2005.full.pdf}
  {http://www.pnas.org/content/110/6/2005.full.pdf} \BibitemShut {NoStop}%
\bibitem [{\citenamefont {Pollakis}\ \emph {et~al.}(2014)\citenamefont
  {Pollakis}, \citenamefont {Wetzel}, \citenamefont {J\"org}, \citenamefont
  {Rave}, \citenamefont {Fettweis},\ and\ \citenamefont
  {J\"ulicher}}]{Pollakis2014}%
  \BibitemOpen
  \bibfield  {author} {\bibinfo {author} {\bibfnamefont {A.}~\bibnamefont
  {Pollakis}}, \bibinfo {author} {\bibfnamefont {L.}~\bibnamefont {Wetzel}},
  \bibinfo {author} {\bibfnamefont {D.~J.}\ \bibnamefont {J\"org}}, \bibinfo
  {author} {\bibfnamefont {W.}~\bibnamefont {Rave}}, \bibinfo {author}
  {\bibfnamefont {G.}~\bibnamefont {Fettweis}},\ and\ \bibinfo {author}
  {\bibfnamefont {F.}~\bibnamefont {J\"ulicher}},\ }\bibfield  {title}
  {\enquote {\bibinfo {title} {Synchronization in networks of mutually
  delay-coupled phase-locked loops},}\ }\href
  {http://stacks.iop.org/1367-2630/16/i=11/a=113009} {\bibfield  {journal}
  {\bibinfo  {journal} {New Journal of Physics}\ }\textbf {\bibinfo {volume}
  {16}},\ \bibinfo {pages} {113009} (\bibinfo {year} {2014})}\BibitemShut
  {NoStop}%
\bibitem [{\citenamefont {Pikovsky}, \citenamefont {Rosenblum},\ and\
  \citenamefont {Kurths}(2001)}]{pikov01}%
  \BibitemOpen
  \bibfield  {author} {\bibinfo {author} {\bibfnamefont {A.}~\bibnamefont
  {Pikovsky}}, \bibinfo {author} {\bibfnamefont {M.}~\bibnamefont
  {Rosenblum}},\ and\ \bibinfo {author} {\bibfnamefont {J.}~\bibnamefont
  {Kurths}},\ }\href@noop {} {\emph {\bibinfo {title} {Synchronization: A
  Universal Concept in Nonlinear Sciences}}}\ (\bibinfo  {publisher} {Cambridge
  University Press},\ \bibinfo {year} {2001})\BibitemShut {NoStop}%
\bibitem [{\citenamefont {Kuramoto}(2003)}]{kura03}%
  \BibitemOpen
  \bibfield  {author} {\bibinfo {author} {\bibfnamefont {Y.}~\bibnamefont
  {Kuramoto}},\ }\href@noop {} {\emph {\bibinfo {title} {Chemical Oscillations,
  Waves, and Turbulence}}}\ (\bibinfo  {publisher} {Springer-Verlag, Berlin},\
  \bibinfo {year} {2003})\BibitemShut {NoStop}%
\bibitem [{\citenamefont {Nakao}(2016)}]{Nakao2016}%
  \BibitemOpen
  \bibfield  {author} {\bibinfo {author} {\bibfnamefont {H.}~\bibnamefont
  {Nakao}},\ }\bibfield  {title} {\enquote {\bibinfo {title} {Phase reduction
  approach to synchronisation of nonlinear oscillators},}\ }\href
  {https://doi.org/10.1080/00107514.2015.1094987} {\bibfield  {journal}
  {\bibinfo  {journal} {Contemporary Physics}\ }\textbf {\bibinfo {volume}
  {57}},\ \bibinfo {pages} {188--214} (\bibinfo {year} {2016})},\ \Eprint
  {https://arxiv.org/abs/https://doi.org/10.1080/00107514.2015.1094987}
  {https://doi.org/10.1080/00107514.2015.1094987} \BibitemShut {NoStop}%
\bibitem [{\citenamefont {Novi\v{c}enko}\ and\ \citenamefont
  {Pyragas}(2012)}]{physd12}%
  \BibitemOpen
  \bibfield  {author} {\bibinfo {author} {\bibfnamefont {V.}~\bibnamefont
  {Novi\v{c}enko}}\ and\ \bibinfo {author} {\bibfnamefont {K.}~\bibnamefont
  {Pyragas}},\ }\bibfield  {title} {\enquote {\bibinfo {title} {Phase reduction
  of weakly perturbed limit cycle oscillations in time-delay systems},}\
  }\href@noop {} {\bibfield  {journal} {\bibinfo  {journal} {Physica D:
  Nonlinear Phenomena}\ }\textbf {\bibinfo {volume} {241}},\ \bibinfo {pages}
  {1090 -- 1098} (\bibinfo {year} {2012})}\BibitemShut {NoStop}%
\bibitem [{\citenamefont {Kotani}\ \emph {et~al.}(2012)\citenamefont {Kotani},
  \citenamefont {Yamaguchi}, \citenamefont {Ogawa}, \citenamefont {Jimbo},
  \citenamefont {Nakao},\ and\ \citenamefont {Ermentrout}}]{kot12}%
  \BibitemOpen
  \bibfield  {author} {\bibinfo {author} {\bibfnamefont {K.}~\bibnamefont
  {Kotani}}, \bibinfo {author} {\bibfnamefont {I.}~\bibnamefont {Yamaguchi}},
  \bibinfo {author} {\bibfnamefont {Y.}~\bibnamefont {Ogawa}}, \bibinfo
  {author} {\bibfnamefont {Y.}~\bibnamefont {Jimbo}}, \bibinfo {author}
  {\bibfnamefont {H.}~\bibnamefont {Nakao}},\ and\ \bibinfo {author}
  {\bibfnamefont {G.~B.}\ \bibnamefont {Ermentrout}},\ }\bibfield  {title}
  {\enquote {\bibinfo {title} {Adjoint method provides phase response functions
  for delay-induced oscillations},}\ }\href@noop {} {\bibfield  {journal}
  {\bibinfo  {journal} {Phys. Rev. Lett.}\ }\textbf {\bibinfo {volume} {109}},\
  \bibinfo {pages} {044101} (\bibinfo {year} {2012})}\BibitemShut {NoStop}%
\bibitem [{\citenamefont {Novi\v{c}enko}(2015)}]{Novicenko2015}%
  \BibitemOpen
  \bibfield  {author} {\bibinfo {author} {\bibfnamefont {V.}~\bibnamefont
  {Novi\v{c}enko}},\ }\bibfield  {title} {\enquote {\bibinfo {title} {Delayed
  feedback control of synchronization in weakly coupled oscillator networks},}\
  }\href {https://doi.org/10.1103/PhysRevE.92.022919} {\bibfield  {journal}
  {\bibinfo  {journal} {Phys. Rev. E}\ }\textbf {\bibinfo {volume} {92}},\
  \bibinfo {pages} {022919} (\bibinfo {year} {2015})}\BibitemShut {NoStop}%
\bibitem [{\citenamefont {Novi\v{c}enko}\ and\ \citenamefont
  {Ratas}(2018)}]{Novicenko2018}%
  \BibitemOpen
  \bibfield  {author} {\bibinfo {author} {\bibfnamefont {V.}~\bibnamefont
  {Novi\v{c}enko}}\ and\ \bibinfo {author} {\bibfnamefont {I.}~\bibnamefont
  {Ratas}},\ }\bibfield  {title} {\enquote {\bibinfo {title} {In-phase
  synchronization in complex oscillator networks by adaptive delayed feedback
  control},}\ }\href {https://doi.org/10.1103/PhysRevE.98.042302} {\bibfield
  {journal} {\bibinfo  {journal} {Phys. Rev. E}\ }\textbf {\bibinfo {volume}
  {98}},\ \bibinfo {pages} {042302} (\bibinfo {year} {2018})}\BibitemShut
  {NoStop}%
\bibitem [{\citenamefont {Hooton}\ and\ \citenamefont {Amann}(2012)}]{hoo12}%
  \BibitemOpen
  \bibfield  {author} {\bibinfo {author} {\bibfnamefont {E.~W.}\ \bibnamefont
  {Hooton}}\ and\ \bibinfo {author} {\bibfnamefont {A.}~\bibnamefont {Amann}},\
  }\bibfield  {title} {\enquote {\bibinfo {title} {Analytical limitation for
  time-delayed feedback control in autonomous systems},}\ }\href@noop {}
  {\bibfield  {journal} {\bibinfo  {journal} {Phys. Rev. Lett.}\ }\textbf
  {\bibinfo {volume} {109}},\ \bibinfo {pages} {154101} (\bibinfo {year}
  {2012})}\BibitemShut {NoStop}%
\bibitem [{\citenamefont {Burd}(2007)}]{burd07}%
  \BibitemOpen
  \bibfield  {author} {\bibinfo {author} {\bibfnamefont {V.}~\bibnamefont
  {Burd}},\ }\href@noop {} {\emph {\bibinfo {title} {Method of Averaging for
  Differential Equations on an Infinite Interval}}}\ (\bibinfo  {publisher}
  {Taylor \& Francis Group},\ \bibinfo {year} {2007})\BibitemShut {NoStop}%
\bibitem [{\citenamefont {Pyragas}(2001)}]{pyr01}%
  \BibitemOpen
  \bibfield  {author} {\bibinfo {author} {\bibfnamefont {K.}~\bibnamefont
  {Pyragas}},\ }\bibfield  {title} {\enquote {\bibinfo {title} {Control of
  chaos via an unstable delayed feedback controller},}\ }\href@noop {}
  {\bibfield  {journal} {\bibinfo  {journal} {Phys. Rev. Lett.}\ }\textbf
  {\bibinfo {volume} {86}},\ \bibinfo {pages} {2265--2268} (\bibinfo {year}
  {2001})}\BibitemShut {NoStop}%
\bibitem [{\citenamefont {Pyragas}\ and\ \citenamefont
  {Novi\v{c}enko}(2013)}]{Pyragas2013}%
  \BibitemOpen
  \bibfield  {author} {\bibinfo {author} {\bibfnamefont {K.}~\bibnamefont
  {Pyragas}}\ and\ \bibinfo {author} {\bibfnamefont {V.}~\bibnamefont
  {Novi\v{c}enko}},\ }\bibfield  {title} {\enquote {\bibinfo {title}
  {Time-delayed feedback control design beyond the odd-number limitation},}\
  }\href {https://doi.org/10.1103/PhysRevE.88.012903} {\bibfield  {journal}
  {\bibinfo  {journal} {Phys. Rev. E}\ }\textbf {\bibinfo {volume} {88}},\
  \bibinfo {pages} {012903} (\bibinfo {year} {2013})}\BibitemShut {NoStop}%
\bibitem [{\citenamefont {Pyragas}(2002)}]{pyr02}%
  \BibitemOpen
  \bibfield  {author} {\bibinfo {author} {\bibfnamefont {K.}~\bibnamefont
  {Pyragas}},\ }\bibfield  {title} {\enquote {\bibinfo {title} {Analytical
  properties and optimization of time-delayed feedback control},}\ }\href@noop
  {} {\bibfield  {journal} {\bibinfo  {journal} {Phys. Rev. E}\ }\textbf
  {\bibinfo {volume} {66}},\ \bibinfo {pages} {026207} (\bibinfo {year}
  {2002})}\BibitemShut {NoStop}%
\bibitem [{\citenamefont {Fiedler}\ \emph {et~al.}(2010)\citenamefont
  {Fiedler}, \citenamefont {Flunkert}, \citenamefont {H\"{o}vel},\ and\
  \citenamefont {Sch\"{o}ll}}]{Fiedler2010}%
  \BibitemOpen
  \bibfield  {author} {\bibinfo {author} {\bibfnamefont {B.}~\bibnamefont
  {Fiedler}}, \bibinfo {author} {\bibfnamefont {V.}~\bibnamefont {Flunkert}},
  \bibinfo {author} {\bibfnamefont {P.}~\bibnamefont {H\"{o}vel}},\ and\
  \bibinfo {author} {\bibfnamefont {E.}~\bibnamefont {Sch\"{o}ll}},\ }\bibfield
   {title} {\enquote {\bibinfo {title} {Delay stabilization of periodic orbits
  in coupled oscillator systems},}\ }\href
  {https://doi.org/10.1098/rsta.2009.0232} {\bibfield  {journal} {\bibinfo
  {journal} {Philosophical Transactions of the Royal Society A: Mathematical,
  Physical and Engineering Sciences}\ }\textbf {\bibinfo {volume} {368}},\
  \bibinfo {pages} {319--341} (\bibinfo {year} {2010})},\ \Eprint
  {https://arxiv.org/abs/https://royalsocietypublishing.org/doi/pdf/10.1098/rsta.2009.0232}
  {https://royalsocietypublishing.org/doi/pdf/10.1098/rsta.2009.0232}
  \BibitemShut {NoStop}%
\bibitem [{\citenamefont {Schneider}(2013)}]{Schneider2013}%
  \BibitemOpen
  \bibfield  {author} {\bibinfo {author} {\bibfnamefont {I.}~\bibnamefont
  {Schneider}},\ }\bibfield  {title} {\enquote {\bibinfo {title} {Delayed
  feedback control of three diffusively coupled {S}tuart-{L}andau oscillators:
  a case study in equivariant {H}opf bifurcation},}\ }\href
  {https://doi.org/10.1098/rsta.2012.0472} {\bibfield  {journal} {\bibinfo
  {journal} {Philosophical Transactions of the Royal Society A: Mathematical,
  Physical and Engineering Sciences}\ }\textbf {\bibinfo {volume} {371}},\
  \bibinfo {pages} {20120472} (\bibinfo {year} {2013})},\ \Eprint
  {https://arxiv.org/abs/https://royalsocietypublishing.org/doi/pdf/10.1098/rsta.2012.0472}
  {https://royalsocietypublishing.org/doi/pdf/10.1098/rsta.2012.0472}
  \BibitemShut {NoStop}%
\bibitem [{\citenamefont {Schneider}\ and\ \citenamefont
  {Bosewitz}(2016)}]{Schneider2016}%
  \BibitemOpen
  \bibfield  {author} {\bibinfo {author} {\bibfnamefont {I.}~\bibnamefont
  {Schneider}}\ and\ \bibinfo {author} {\bibfnamefont {M.}~\bibnamefont
  {Bosewitz}},\ }\bibfield  {title} {\enquote {\bibinfo {title} {Eliminating
  restrictions of time-delayed feedback control using equivariance},}\ }\href
  {https://doi.org/10.3934/dcds.2016.36.451} {\bibfield  {journal} {\bibinfo
  {journal} {Discrete \& Continuous Dynamical Systems - A}\ }\textbf {\bibinfo
  {volume} {36}},\ \bibinfo {pages} {451} (\bibinfo {year} {2016})}\BibitemShut
  {NoStop}%
\bibitem [{\citenamefont {Kiss}(2018)}]{Kiss2018}%
  \BibitemOpen
  \bibfield  {author} {\bibinfo {author} {\bibfnamefont {I.~Z.}\ \bibnamefont
  {Kiss}},\ }\bibfield  {title} {\enquote {\bibinfo {title} {Synchronization
  engineering},}\ }\href
  {https://doi.org/https://doi.org/10.1016/j.coche.2018.02.006} {\bibfield
  {journal} {\bibinfo  {journal} {Current Opinion in Chemical Engineering}\
  }\textbf {\bibinfo {volume} {21}},\ \bibinfo {pages} {1--9} (\bibinfo {year}
  {2018})}\BibitemShut {NoStop}%
\bibitem [{\citenamefont {Kiss}\ \emph {et~al.}(2007)\citenamefont {Kiss},
  \citenamefont {Rusin}, \citenamefont {Kori},\ and\ \citenamefont
  {Hudson}}]{Kiss2007}%
  \BibitemOpen
  \bibfield  {author} {\bibinfo {author} {\bibfnamefont {I.~Z.}\ \bibnamefont
  {Kiss}}, \bibinfo {author} {\bibfnamefont {C.~G.}\ \bibnamefont {Rusin}},
  \bibinfo {author} {\bibfnamefont {H.}~\bibnamefont {Kori}},\ and\ \bibinfo
  {author} {\bibfnamefont {J.~L.}\ \bibnamefont {Hudson}},\ }\bibfield  {title}
  {\enquote {\bibinfo {title} {Engineering complex dynamical structures:
  Sequential patterns and desynchronization},}\ }\href
  {https://doi.org/10.1126/science.1140858} {\bibfield  {journal} {\bibinfo
  {journal} {Science}\ }\textbf {\bibinfo {volume} {316}},\ \bibinfo {pages}
  {1886--1889} (\bibinfo {year} {2007})}\BibitemShut {NoStop}%
\bibitem [{\citenamefont {Rusin}\ \emph {et~al.}(2009)\citenamefont {Rusin},
  \citenamefont {Kiss}, \citenamefont {Kori},\ and\ \citenamefont
  {Hudson}}]{Rusin2009}%
  \BibitemOpen
  \bibfield  {author} {\bibinfo {author} {\bibfnamefont {C.~G.}\ \bibnamefont
  {Rusin}}, \bibinfo {author} {\bibfnamefont {I.~Z.}\ \bibnamefont {Kiss}},
  \bibinfo {author} {\bibfnamefont {H.}~\bibnamefont {Kori}},\ and\ \bibinfo
  {author} {\bibfnamefont {J.~L.}\ \bibnamefont {Hudson}},\ }\bibfield  {title}
  {\enquote {\bibinfo {title} {Framework for engineering the collective
  behavior of complex rhythmic systems},}\ }\href
  {https://doi.org/10.1021/ie801807f} {\bibfield  {journal} {\bibinfo
  {journal} {Industrial {\&} Engineering Chemistry Research}\ }\textbf
  {\bibinfo {volume} {48}},\ \bibinfo {pages} {9416--9422} (\bibinfo {year}
  {2009})}\BibitemShut {NoStop}%
\bibitem [{\citenamefont {Pyragas}\ and\ \citenamefont
  {Pyragas}(2011)}]{Pyragas2011}%
  \BibitemOpen
  \bibfield  {author} {\bibinfo {author} {\bibfnamefont {V.}~\bibnamefont
  {Pyragas}}\ and\ \bibinfo {author} {\bibfnamefont {K.}~\bibnamefont
  {Pyragas}},\ }\bibfield  {title} {\enquote {\bibinfo {title} {Adaptive
  modification of the delayed feedback control algorithm with a continuously
  varying time delay},}\ }\href
  {https://doi.org/https://doi.org/10.1016/j.physleta.2011.08.072} {\bibfield
  {journal} {\bibinfo  {journal} {Physics Letters A}\ }\textbf {\bibinfo
  {volume} {375}},\ \bibinfo {pages} {3866 -- 3871} (\bibinfo {year}
  {2011})}\BibitemShut {NoStop}%
\bibitem [{\citenamefont {Novi\v{c}enko}()}]{novi2020}%
  \BibitemOpen
  \bibfield  {author} {\bibinfo {author} {\bibfnamefont {V.}~\bibnamefont
  {Novi\v{c}enko}},\ }\href@noop {} {\enquote {\bibinfo {title} {Comment on
  "{A}daptive modification of the delayed feedback control algorithm with a
  continuously varying time delay"
  https://doi.org/10.1016/j.physleta.2011.08.072},}\ }\Eprint
  {https://arxiv.org/abs/2003.01596} {arXiv:2003.01596} \BibitemShut {NoStop}%
\bibitem [{\citenamefont {Tama\v{s}evi\v{c}ius}\ \emph
  {et~al.}(2008)\citenamefont {Tama\v{s}evi\v{c}ius}, \citenamefont
  {Tama\v{s}evi\v{c}i\={u}t\.{e}}, \citenamefont {Mykolaitis},\ and\
  \citenamefont {Bumelien\.{e}}}]{tamasevicius08}%
  \BibitemOpen
  \bibfield  {author} {\bibinfo {author} {\bibfnamefont {A.}~\bibnamefont
  {Tama\v{s}evi\v{c}ius}}, \bibinfo {author} {\bibfnamefont {E.}~\bibnamefont
  {Tama\v{s}evi\v{c}i\={u}t\.{e}}}, \bibinfo {author} {\bibfnamefont
  {G.}~\bibnamefont {Mykolaitis}},\ and\ \bibinfo {author} {\bibfnamefont
  {S.}~\bibnamefont {Bumelien\.{e}}},\ }\bibfield  {title} {\enquote {\bibinfo
  {title} {Switching from stable to unknown unstable steady states of dynamical
  systems},}\ }\href {https://doi.org/10.1103/PhysRevE.78.026205} {\bibfield
  {journal} {\bibinfo  {journal} {Phys. Rev. E}\ }\textbf {\bibinfo {volume}
  {78}},\ \bibinfo {pages} {026205} (\bibinfo {year} {2008})}\BibitemShut
  {NoStop}%
\bibitem [{\citenamefont {Tama\v{s}evi\v{c}ius}\ \emph
  {et~al.}(2007)\citenamefont {Tama\v{s}evi\v{c}ius}, \citenamefont
  {Mykolaitis}, \citenamefont {Pyragas},\ and\ \citenamefont
  {Pyragas}}]{tam07}%
  \BibitemOpen
  \bibfield  {author} {\bibinfo {author} {\bibfnamefont {A.}~\bibnamefont
  {Tama\v{s}evi\v{c}ius}}, \bibinfo {author} {\bibfnamefont {G.}~\bibnamefont
  {Mykolaitis}}, \bibinfo {author} {\bibfnamefont {V.}~\bibnamefont
  {Pyragas}},\ and\ \bibinfo {author} {\bibfnamefont {K.}~\bibnamefont
  {Pyragas}},\ }\bibfield  {title} {\enquote {\bibinfo {title} {Delayed
  feedback control of periodic orbits without torsion in nonautonomous chaotic
  systems: Theory and experiment},}\ }\href@noop {} {\bibfield  {journal}
  {\bibinfo  {journal} {Phys. Rev. E}\ }\textbf {\bibinfo {volume} {76}},\
  \bibinfo {pages} {026203} (\bibinfo {year} {2007})}\BibitemShut {NoStop}%
\bibitem [{\citenamefont {Goldsztein}, \citenamefont {Nadeau},\ and\
  \citenamefont {Strogatz}(2020)}]{goldsztein2020antiphase}%
  \BibitemOpen
  \bibfield  {author} {\bibinfo {author} {\bibfnamefont {G.~H.}\ \bibnamefont
  {Goldsztein}}, \bibinfo {author} {\bibfnamefont {A.~N.}\ \bibnamefont
  {Nadeau}},\ and\ \bibinfo {author} {\bibfnamefont {S.~H.}\ \bibnamefont
  {Strogatz}},\ }\href@noop {} {\enquote {\bibinfo {title} {Antiphase versus
  in-phase synchronization of coupled pendulum clocks and metronomes},}\ }
  (\bibinfo {year} {2020}),\ \Eprint {https://arxiv.org/abs/2008.02947}
  {arXiv:2008.02947 [nlin.AO]} \BibitemShut {NoStop}%
\bibitem [{\citenamefont {Pantaleone}(2002)}]{Pantaleone2002}%
  \BibitemOpen
  \bibfield  {author} {\bibinfo {author} {\bibfnamefont {J.}~\bibnamefont
  {Pantaleone}},\ }\bibfield  {title} {\enquote {\bibinfo {title}
  {Synchronization of metronomes},}\ }\href {https://doi.org/10.1119/1.1501118}
  {\bibfield  {journal} {\bibinfo  {journal} {American Journal of Physics}\
  }\textbf {\bibinfo {volume} {70}},\ \bibinfo {pages} {992--1000} (\bibinfo
  {year} {2002})},\ \Eprint
  {https://arxiv.org/abs/https://doi.org/10.1119/1.1501118}
  {https://doi.org/10.1119/1.1501118} \BibitemShut {NoStop}%
\bibitem [{\citenamefont {Chicone}(2006)}]{Chicone2006}%
  \BibitemOpen
  \bibfield  {author} {\bibinfo {author} {\bibfnamefont {C.}~\bibnamefont
  {Chicone}},\ }\enquote {\bibinfo {title} {Linear systems and stability of
  nonlinear systems},}\ in\ \href {https://doi.org/10.1007/0-387-35794-7_2}
  {\emph {\bibinfo {booktitle} {Ordinary Differential Equations with
  Applications}}}\ (\bibinfo  {publisher} {Springer New York},\ \bibinfo
  {address} {New York, NY},\ \bibinfo {year} {2006})\ pp.\ \bibinfo {pages}
  {145--224}\BibitemShut {NoStop}%
\end{thebibliography}%

\end{document}